\documentclass[modern]{aastex62}

\usepackage{graphicx, color}
\usepackage{dcolumn}
\usepackage{bm}
\usepackage{booktabs}
\usepackage{natbib}
\usepackage{amsmath}
\usepackage{threeparttable}

\newcommand {\vect}[1]{\mbox{\boldmath $#1$}}

\newcommand {\dif}[3][]{\frac{d^{#1}#2}{d#3^{#1}}}
\newcommand {\pdif}[3][]{\frac{\partial^{#1}#2}{\partial#3^{#1}}}

\makeatletter
\def\mart{\@ifnextchar[{\mart@@}{\mart@}}
\def\mart@@[#1]#2{\sqrt[#1]{\mathstrut{#2}}}
\def\mart@#1{\sqrt{\mathstrut{#1}}}
\makeatother
\newcommand {\Alfven}{Alfv\'{e}n}
\newcommand {\Alfvenic}{Alfv\'{e}nic}

\newcommand{\myemail}{minoshim@jamstec.go.jp}

\newcommand{\sgn}{{\rm sgn}}

\long\def\symbolfootnote[#1]#2{\begingroup%
\def\thefootnote{\fnsymbol{footnote}}\footnote[#1]{#2}\endgroup}

\begin{document}

\title{A Multistate Low-dissipation Advection Upstream Splitting Method for Ideal Magnetohydrodynamics}
\shorttitle{Multistate Low-dissipation AUSM for MHD}

\author{Takashi Minoshima}
\affiliation{Center for Mathematical Science and Advanced Technology, Japan Agency for Marine-Earth Science and Technology, 3173-25, Syowa-machi, Kanazawaku, Yokohama 236-0001, Japan}

\author{Keiichi Kitamura}
\affiliation{Yokohama National University, 79-5 Tokiwadai, Hodogaya-ku, Yokohama, Kanagawa 240-8501, Japan}

\author{Takahiro Miyoshi}
\affiliation{Graduate School of Advanced Science and Engineering, Hiroshima University, 1-3-1, Kagamiyama, Higashi-hiroshima, 739-8526, Japan}

\correspondingauthor{Takashi Minoshima}
\shortauthors{Minoshima et al.}
\email{\myemail}

\begin{abstract}
We develop a new numerical scheme for ideal magnetohydrodynamic (MHD) simulations, which is robust against one- and multi-dimensional shocks, and is accurate for low Mach number flows and discontinuities.
The scheme belongs to a family of the advection upstream splitting method employed in computational aerodynamics, and it splits the inviscid flux in MHD equations into advection, pressure, and magnetic tension parts, and then individually evaluates mass, pressure, and magnetic tension fluxes at the interface of a computational cell.
The mass flux is designed to avoid numerical shock instability in multidimension, while preserving contact discontinuity.
The pressure flux possesses a proper scaling for low Mach number flows, allowing reliable simulations of nearly incompressible flows.
The magnetic tension flux is built to be consistent with the HLLD approximate Riemann solver to preserve rotational discontinuity.
We demonstrate various benchmark tests to verify the novel performance of the scheme.
Our results indicate that the scheme must be a promising tool to tackle astrophysical systems that include both low and high Mach number flows, as well as magnetic field inhomogeneities.
\end{abstract}

\keywords{Magnetohydrodynamical simulations (1966), Magnetic fields (994), Shocks (2086)}

\section{Introduction}\label{sec:introduction}
Magnetohydrodynamic (MHD) simulation is an indispensable way to study macroscopic dynamics in space and astrophysical plasmas, which include supersonic flows and shocks, turbulence, amplification and dissipation of magnetic fields, non-ideal effects such as kinetic effects, radiative transfer, and relativistic effects, and so on. 
The rapid increase of computational resources enables further large-scale and long-term numerical simulations.
Numerous studies have been devoted to developing novel numerical techniques to fully utilize an available resource and then reveal its underlying physics in highly nonlinear systems as accurately as possible.
In the simulation of compressible fluids, a family of upwind, shock-capturing schemes have succeeded in resolving shocks and discontinuities since the pioneering work by \cite{1959GODUNOV}.
Many compressible MHD simulation codes employ shock-capturing schemes, such as the Roe-type flux difference splitting (FDS) method \citep[e.g.,][]{1981JCoPh..43..357R,1988JCoPh..75..400B,1998ApJS..116..119B} and the Harten-Lax-van Leer-type (HLL) method \citep{1983SIAMrev.25..35M,2005JCoPh.203..344L,2005JCoPh.208..315M}, as a building block.
Of these, the HLLD approximate Riemann solver developed by \cite{2005JCoPh.208..315M} possesses shock-capturing capability with satisfactory accuracy and stability, and is implemented in various modern MHD simulation codes \citep[e.g.,][]{2006A&A...457..371F,2007ApJS..170..228M,2008ApJS..178..137S,2009JCoPh.228..952L,2011ApJ...737...13K,2011PhPl...18b2105Z,2019PASJ...71...83M}.

Although the shock-capturing scheme has enhanced striking progress in compressible fluid simulation, there remain numerical difficulties in practical simulation studies.
The scheme tends to suffer from numerical instability when a multidimensional shock is well aligned to the grid spacing.
The resulting solution is catastrophic, such as the odd-even decoupling and the Carbuncle phenomena \citep{1994IJNMF..18..555Q}.
This may be problematic when one conducts a simulation including steady high Mach number shocks, e.g., a bow shock ahead of the planetary magnetosphere.
The schemes without preserving contact discontinuity, such as the flux vector splitting method \cite[FVS;][]{1981JCoPh..40..263S}, are known to be less sensitive to the instability, indicating that the dissipation of density gradient is effective for suppressing the instability \citep{2001JCoPh.166..271P}.
Therefore, a straightforward way to cure the instability entails employing a non-contact preserving scheme \citep{1994IJNMF..18..555Q,1997SIAM.18..633K,2008JCoPh.227.2560N} or adding artificial viscosity \citep{2001JCoPh.166..271P,2008JCoPh.227.7952H,2008ApJS..178..137S} at regions dangerous to the instability.

Meanwhile, the shock-capturing scheme is not necessarily appropriate for low Mach number flows because it numerically dissipates the velocity jump with a scale of sound speed (shown later in Section \ref{sec:pressure-flux}).
Owing to excessive numerical diffusion, the solution of low Mach number flows obtained with familiar compressible fluid simulations deviates from a correct solution with decreasing Mach number.
One may encounter this problem in nearly incompressible flow simulations, e.g., dynamo process in stellar convection zones.
To obtain a reliable solution of low Mach number flows, one may consider the infinite speed of sound \cite[incompressible or anelastic approximation. e.g., ][]{2001ApJ...554.1175M,2004ApJ...612..276S,2005A&A...444...25L}, or inversely, artificially reduce the sound speed to relax the stiffness of the hyperbolic system of equations \citep{1995AIAAJ..33.2050W}.

A family of the advection upstream splitting method \cite[AUSM;][]{1993JCoPh.107...23L} has been extensively employed in computational aerodynamics.
The AUSM-family scheme is an alternative to the conventional FVS and FDS methods, for improving the accuracy, stability, and computational efficiency.
In contrast to the FVS and FDS methods, the AUSM-family scheme is free from the calculation of the Jacobian
matrix, which reduces computational cost and makes it easier to extend the scheme to other hyperbolic systems of equations.
Some of them are known to be robust against numerical shock instability \citep{2000JCoPh.160..623L}.
Furthermore, recent AUSM-family schemes are extended to ``all-speed'' regime, allowing reliable simulations of nearly incompressible flows with a compressible code \citep{2006JCoPh.214..137L,2011AIAAJ..49.1693S,2013JCoPh.245...62K}.

Although the above-mentioned properties of the AUSM-family scheme must be quite useful for MHD simulations as well, the extension of the scheme has been limited \citep{2009AIAAJ..47..970H,2012JCoPh.231.6233S,2014JCoPh.275..323X,2019ShWav..29..611K}, and its advantage is not fully utilized in MHD simulations thus far.
Furthermore, preceding studies do not particularly focus on resolving the magnetic tension force; and thus, the quality of their solution of the {\Alfven} wave is not necessarily as high as obtainable in the standard Roe and HLLD schemes.

Consequently, we propose a new AUSM-family scheme for ideal MHD simulations to address the above-mentioned numerical difficulties inherent in familiar shock-capturing schemes, and then tackle astrophysical systems, including both low and high Mach number flows.
As opposed to the preceding AUSM-family schemes, ours is designed to capture MHD tangential and rotational discontinuities as well as contact discontinuity.
Section \ref{sec:ausm-for-hd} briefly describes the AUSM-family scheme for hydrodynamic simulations.
Subsequently, we detail the extension of the scheme to MHD simulations in Section \ref{sec:new-ausm-mhd}.
We employ the AUSM$^+$-up scheme \citep{2006JCoPh.214..137L} and the HLLD scheme as a building block.
Section \ref{sec:numerical-tests} presents numerical simulation results of various benchmark tests, including shocks, waves, and very low and high Mach number flows, to assess the performance of the present scheme.
Finally, Section \ref{sec:summary} summarizes the paper.
The source code (written in the C programming language) for the numerical tests in this paper can be downloaded from the GitHub website\footnote{\href{https://github.com/minoshim/MLAU}{https://github.com/minoshim/MLAU}}.

\section{AUSM-family Scheme for Hydrodynamics}\label{sec:ausm-for-hd}
We begin with one-dimensional Euler equations written in conservative form as follows:
\begin{eqnarray}
 \pdif{\vect{U}}{t} + \pdif{ \vect{F}}{x} = 0,
 \vect{U} = 
\begin{bmatrix}
 \rho \\
 {\rho u} \\
 e \\
\end{bmatrix}
,\vect{F} = 
\begin{bmatrix}
 \rho u \\
 \rho u^2 + P  \\
\rho u h\\
\end{bmatrix}
,\label{eq:2}
\end{eqnarray}
where $\rho,u,e,P$, and $h=(e+P)/\rho$ are the fluid mass density, velocity, total energy density, gas pressure, and total enthalpy, respectively.
The gas pressure is determined from the equation of state, i.e.,
\begin{eqnarray}
 P = \left(\gamma-1\right)\left(e-\frac{\rho u^2}{2}\right),\label{eq:3}
\end{eqnarray}
where $\gamma$ is the specific heat ratio.
We discretize Equation (\ref{eq:2}) into a finite volume form as:
\begin{eqnarray}
\dif{\vect{U}^n_i}{t} = - \frac{\vect{\hat{F}}_{i+1/2}-\vect{\hat{F}}_{i-1/2}}{\Delta x},\vect{U}^n_i=\frac{1}{\Delta x}\int_{x_i-\Delta x/2}^{x_i+\Delta x/2} \vect{U}(x,t_n)dx,\label{eq:4}
\end{eqnarray}
where $\vect{\hat{F}}_{i \pm 1/2}$ is a numerical flux at the interfaces of a cell $I_i=[x_i-\Delta x/2,x_i+\Delta x/2]$.
The evaluation of the numerical flux largely impacts the quality of numerical solutions.

\begin{figure}[htbp]
\centering
\includegraphics[clip,angle=0,scale=0.4]{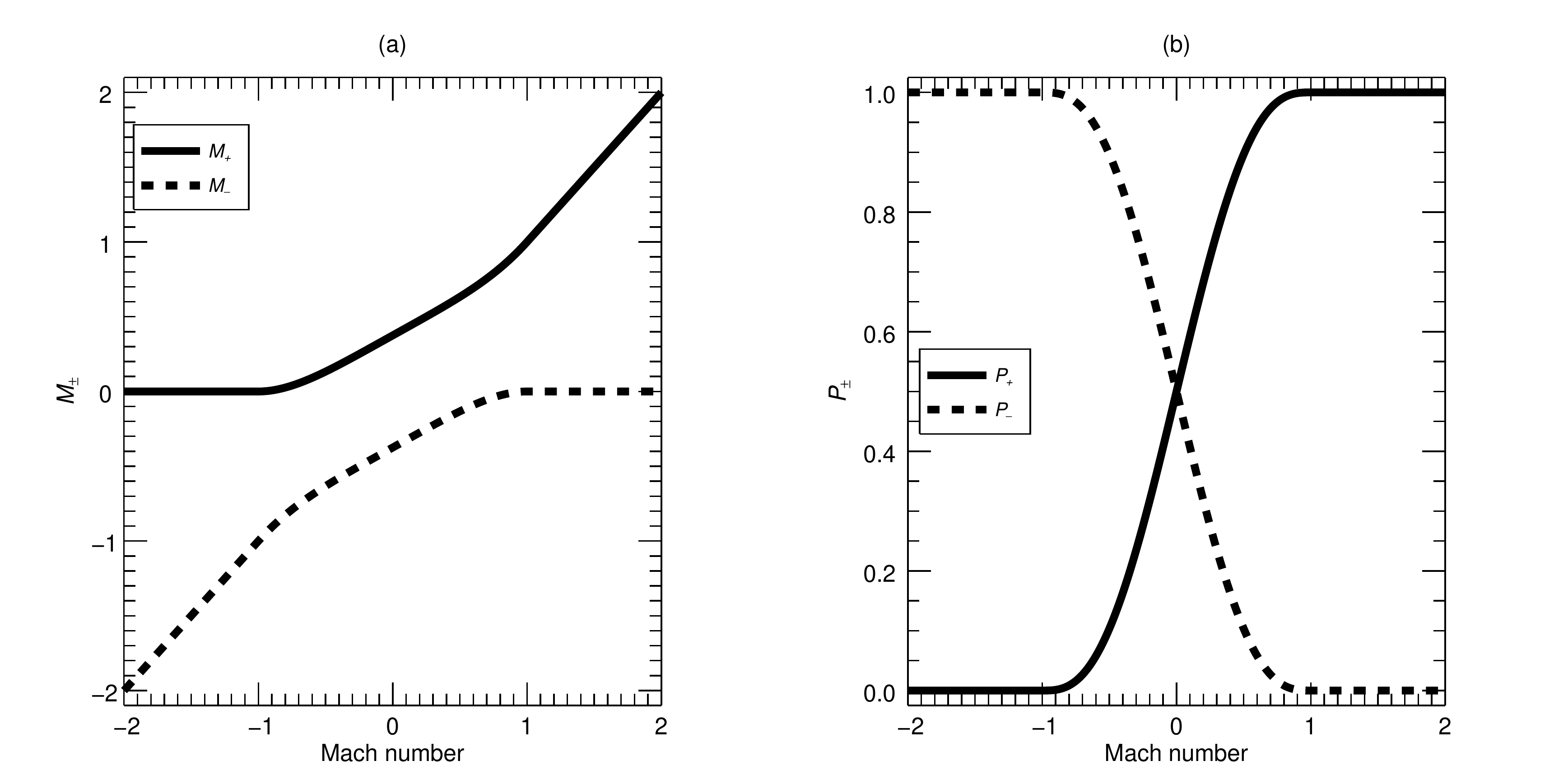}
\caption{Split functions used for (a) the mass flux and (b) the pressure flux of the AUSM$^+$ scheme (Equations (\ref{eq:6}) and (\ref{eq:8})).}
\label{fig:ausm_func} 
\end{figure}

\cite{1993JCoPh.107...23L} proposed the advection upstream splitting method (AUSM) to calculate a simple, yet accurate and robust numerical flux.
The scheme splits the numerical flux into advection and pressure parts, and determines them individually:
\begin{eqnarray}
&& \vect{\hat{F}}_{i+1/2}=\dot{m}\left(d_L\vect{\Phi}_{i+1/2,L}+d_R\vect{\Phi}_{i+1/2,R}\right) + \hat{P}_{i+1/2}\vect{N},\label{eq:52}\\
&& d_L=\frac{1+\sgn(\dot{m})}{2},d_R=\frac{1-\sgn(\dot{m})}{2},\vect{\Phi}=\left(1,u,h\right)^T, \vect{N}=\left(0,1,0\right)^T,\label{eq:5}
\end{eqnarray}
where the subscripts $L,R$ refer to the left and right states at the interface, and $\dot{m}=(\rho u)_{i+1/2}$ and $\hat{P}_{i+1/2}$ are the interface mass and pressure fluxes, respectively,  evaluated from the left- and right-side variables $\vect{U}_{i+1/2,L},\vect{U}_{i+1/2,R}.$
The advection part (first term of Equation (\ref{eq:52})) is upwinded with respect to the interface mass flux. 
Various AUSM-family schemes have been proposed to calculate better mass and pressure fluxes.
For example, \cite{1996JCoPh.129..364L} proposed the AUSM$^+$ scheme to preserve the contact discontinuity that is smeared by the early AUSM scheme:
\begin{eqnarray}
&& \dot{m} = M_{i+1/2}a_{i+1/2}\left(\frac{1+\sgn(M_{i+1/2})}{2}\rho_{i+1/2,L} + \frac{1-\sgn(M_{i+1/2})}{2}\rho_{i+1/2,R}\right),\label{eq:1}\\
&& M_{i+1/2} = {\cal M}_{+}(M_{i+1/2,L})+{\cal M}_{-}(M_{i+1/2,R}),M_{i+1/2,\alpha}=\frac{u_{i+1/2,\alpha}}{a_{i+1/2}},(\alpha=L,R),\label{eq:6}\\
&& {\cal M}_{\pm}(M) = \left\{
\begin{array}{l}
 \left(M \pm |M|\right)/2, \;\;\; {\rm if} \; |M|>1\\
 \pm \left(1 \pm M\right)^2/4 \pm \left(1-M^2\right)^2/8, \;\;\; {\rm otherwise}\\
\end{array}
\right.\label{eq:7}\\
&& \hat{P}_{i+1/2} = {\cal P}_{+}(M_{i+1/2,L})P_{i+1/2,L} + {\cal P}_{-}(M_{i+1/2,R})P_{i+1/2,R},\label{eq:8}\\
&& {\cal P}_{\pm}(M) = \left\{
\begin{array}{l}
 \left(1 \pm \sgn(M)\right)/2, \;\;\; {\rm if} \; |M|>1\\
 \left(1 \pm M\right)^2\left(2\mp M\right)/4 \pm 3M\left(1-M^2\right)^2/16, \;\;\; {\rm otherwise}\\
\end{array}
\right.\label{eq:9}
\end{eqnarray}
where $a_{i+1/2}$ and $M_{i+1/2}$ are the interface sound speed and Mach number, respectively.
The interface Mach number and pressure flux are weighted averages of the left and right states (Equations (\ref{eq:6}) and (\ref{eq:8})) by differentiable high-order polynomials (Equations (\ref{eq:7}) and (\ref{eq:9}), shown in Figure \ref{fig:ausm_func}), and the interface mass flux is upwinded with respect to the interface Mach number (Equation (\ref{eq:1})).
\cite{2000JCoPh.160..623L} demonstrated that the AUSM$^+$ scheme suffers less from numerical shock instability in spite of its contact-preserving nature, while other contact-preserving schemes, such as Roe and HLLC \citep{1994ShWav...4...25T}, are vulnerable to it.

\section{Multistate Low-dissipation AUSM for MHD}\label{sec:new-ausm-mhd}
Now we move on to ideal MHD equations.
The state variables and the corresponding inviscid fluxes are as follows:
\begin{eqnarray}
\vect{U} = 
\begin{bmatrix}
 \rho \\
 {\rho u} \\
 {\rho v} \\
 {\rho w} \\
 {B_y} \\
 {B_z} \\
 e \\
\end{bmatrix}
, \vect{F} = 
\begin{bmatrix}
 \rho u \\
 \rho u^2 + P + \left(B_y^2+B_z^2-B_x^2\right)/2 \\
 \rho v u - B_x B_y \\
 \rho w u - B_x B_z \\
 B_y u - B_x v \\
 B_z u - B_x w \\
\left(e+P+|\vect{B}|^2/2\right)u - B_x\left(\vect{u}\cdot\vect{B}\right)\\
\end{bmatrix}
, P = \left(\gamma-1\right)\left(e-\frac{\rho |\vect{u}|^2}{2} - \frac{|\vect{B}|^2}{2}\right),\nonumber \\
 \label{eq:10}
\end{eqnarray}
where $\vect{u}=(u,v,w)$ and $\vect{B}=(B_x,B_y,B_z)$ is the magnetic field.
The divergence-free condition for the magnetic field gives $B_x = {\rm Constant}$ in one dimension.
To numerically solve the MHD equations by the AUSM-family scheme, we propose to split the flux into three parts, namely, advection, pressure, and magnetic tension parts:
\begin{eqnarray}
\vect{F} = \rho u \vect{\Phi} + P_t \vect{N} - \vect{T}
,\label{eq:76}
\end{eqnarray}
where,
\begin{eqnarray}
&& \vect{\Phi}=\left(1,u,v,w,B_y/\rho,B_z/\rho,h\right)^T,\; \label{eq:12}\\
 && \vect{N}=\left(0,1,0,0,0,0,0\right)^T,\label{eq:13}\\
&&  \vect{T}= B_x\left(0,B_x/2,B_y,B_z,v,w,\vect{u}_t \cdot \vect{B_t}\right)^T,\label{eq:14}\\
&& \vect{u}_t=(0,v,w),\vect{B}_t=(0,B_y,B_z),\label{eq:78}\\
&& h=\gamma P/(\gamma-1)\rho+|\vect{u}|^2/2+|\vect{B}_t|^2/\rho,\label{eq:77}
\end{eqnarray}
and
\begin{eqnarray}
P_t=P+|\vect{B}_t|^2/2,\label{eq:80}  
\end{eqnarray}
is the total pressure (except for the contribution of $B_x$).
Subsequently, we express the numerical flux at the interface in the same manner as Equation (\ref{eq:52}),
\begin{eqnarray}
&& \vect{\hat{F}}_{i+1/2}=\dot{m}\left(d_L\vect{\Phi}_{i+1/2,L}+d_R\vect{\Phi}_{i+1/2,R}\right) + \hat{P}_{t,i+1/2}\vect{N}-\vect{\hat{T}}_{i+1/2},\label{eq:11}\\
&& d_L=\frac{1+\sgn(\dot{m})}{2},d_R=\frac{1-\sgn(\dot{m})}{2},\dot{m}=\left(\rho u\right)_{i+1/2}\label{eq:53}
\end{eqnarray}
where $\vect{\hat{T}}$ is termed the magnetic tension flux.
A critical issue involves the evaluation of the mass, pressure, and magnetic tension fluxes $(\dot{m},\hat{P}_{t,i+1/2},\vect{\hat{T}}_{i+1/2})$ to get reliable solutions.

\subsection{Mass Flux}\label{sec:mass-flux}
We build a mass flux based on the AUSM$^+$-up scheme developed by \cite{2006JCoPh.214..137L}, which adds a pressure difference term to the mass flux of the AUSM$^+$ scheme.
To extend the scheme to MHD, we scale the functions in Equations (\ref{eq:1})-(\ref{eq:6}) by the fast magnetosonic speed $c_f$ rather than the sound speed $a$.
The interface mass flux $\dot{m}$ is expressed as:
\begin{eqnarray}
&& \dot{m} = M_{i+1/2}c_{f,i+1/2}\left(\frac{1+\sgn(M_{i+1/2})}{2}\rho_{i+1/2,L} + \frac{1-\sgn(M_{i+1/2})}{2}\rho_{i+1/2,R}\right),\label{eq:70}\\
&& M_{i+1/2} = M^*_{i+1/2} - \max(1-|M^*_{i+1/2}|,0) \frac{\Delta P_{t,i+1/2}}{(\rho_{i+1/2,L}+\rho_{i+1/2,R})c^2_{f,i+1/2}},\label{eq:16}\\
&& M^*_{i+1/2} = {\cal M}_{+}(M_{i+1/2,L})+{\cal M}_{-}(M_{i+1/2,R}),\;\;\;M_{i+1/2,\alpha}=\frac{u_{i+1/2,\alpha}}{c_{f,i+1/2}},(\alpha=L,R),\label{eq:15}\\
&& c^2_{f} = \frac{1}{2}\left[\left(c^2_a + a^2\right) + \sqrt{\left(c^2_a + a^2\right)^2-4a^2 c^2_{ax}}\right],c^2_a=\frac{\vect{|B|}^2}{\rho},c^2_{ax}=\frac{B^2_x}{\rho},a^2=\frac{\gamma P}{\rho},\label{eq:17}
\end{eqnarray}
where $c_{f,i+1/2}=\max(c_{f,i+1/2,L},c_{f,i+1/2,R})$, $\Delta P_t = P_{t,R}-P_{t,L}$, and the function ${\cal M}_\pm$ is defined in Equation (\ref{eq:7}).

\cite{2000JCoPh.160..623L} suggested that the pressure difference term in the mass flux is a possible cause of numerical shock instability in multidimension, although it is effective for stabilizing one-dimensional shocks.
\cite{2003JCoPh.185..342K} argued that the pressure difference term converts the pressure perturbation into the density perturbation, and the latter along the shock surface is not dissipated with the contact-preserving scheme when the shock is rested in the direction parallel to the shock.
If the pressure perturbation is continuously injected, it becomes unstable.
To avoid the instability but maintain the robustness of one-dimensional shocks, we eliminate the pressure difference term only at regions dangerous to the instability.
Following \cite{2013AIAAJ..51..992S}, the pressure difference term in Equation (\ref{eq:16}) is multiplied by a shock-detecting factor $\theta$ as a function of the velocity difference:
\begin{eqnarray}
\Delta P_t &\leftarrow& \theta \Delta P_t,\label{eq:51}\\
\theta &=& {\rm min}\left(1,\frac{-{\rm min}(\Delta u,0)+{c}_f}{-{\rm min}(\Delta v, \Delta w, 0)+{c}_f}\right)^4,\label{eq:47}\\
\Delta u_{i+1/2,j,k} &=& u_{i+1,j,k}-u_{i,j,k},\label{eq:48}\\
\Delta v_{i+1/2,j,k} &=& {\rm min}(v_{i,j,k}-v_{i,j-1,k},v_{i,j+1,k}-v_{i,j,k},\nonumber \\
&& v_{i+1,j,k}-v_{i+1,j-1,k},v_{i+1,j+1,k}-v_{i+1,j,k}),\label{eq:49}\\
\Delta w_{i+1/2,j,k} &=& {\rm min}(w_{i,j,k}-w_{i,j,k-1},w_{i,j,k+1}-w_{i,j,k},\nonumber \\
&& w_{i+1,j,k}-w_{i+1,j,k-1},w_{i+1,j,k+1}-w_{i+1,j,k}).\label{eq:50}
\end{eqnarray}
The factor $\theta$ approaches zero only at a strong shock, whose normal direction is orthogonal to $x$, and it does not violate the preservation of the contact discontinuity.

To compare the present scheme (Equation (\ref{eq:70})) with others, we express the mass flux of upwind schemes for hydrodynamic simulations in a general form as:
\begin{eqnarray}
 \dot{m} = \frac{1}{2}\left[(\rho u)_L + (\rho u)_R - D^{(\rho)}\Delta \rho - D^{(u)} \Delta u - D^{(P)} \Delta P \right],\label{eq:71}
\end{eqnarray}
where $D^{(\rho)},D^{(u)},D^{(P)}$ represent dissipation coefficients associated with the difference in primitive variables \citep{2000JCoPh.160..623L}.
Table \ref{tab:massflux} lists the coefficients at subsonic range among various schemes; AUSM$^+$, AUSM-family schemes with a Roe-type mass flux \citep[SHUS and SLAU;][]{2011AIAAJ..49.1693S}, HLL and HLLC approximate Riemann solvers, and the present one herein.
Although these schemes are built based on different ideas, they possess similar properties, including (i) the density dissipation scales with $|u|$ to preserve the contact discontinuity (except for HLL), (ii) the pressure dissipation works for low Mach number flows (except for AUSM$^+$ and HLL), and (iii) the velocity dissipation coefficient is set such that the mass flux reduces to a one-side value at $|M|=1$ (except for SLAU).
The mass flux of the Roe scheme for MHD simulations has dissipation terms of $\vect{u}_t$ and $\vect{B}_t$ due to the rotation of the tangential components \cite[e.g.,][]{1988JCoPh..75..400B}.
On the other hand, the mass flux of the HLLD scheme omits them and treats the total pressure instead; thus, it is essentially the same as that of the HLLC scheme.
The mass flux of the present scheme follows the strategy of the HLLD scheme.

\subsection{Magnetic Tension Flux}\label{sec:magn-tens-flux}
The preservation of the rotational discontinuity is of critical importance for practical MHD simulations with an aim to solve the evolution of the magnetic field, although preceding AUSM-family schemes do not necessarily pay attention to it.
To accurately solve the rotational discontinuity, we build a magnetic tension flux that is consistent with the HLLD scheme.

\begin{figure}[htbp]
\centering
\includegraphics[clip,angle=0,scale=1.0]{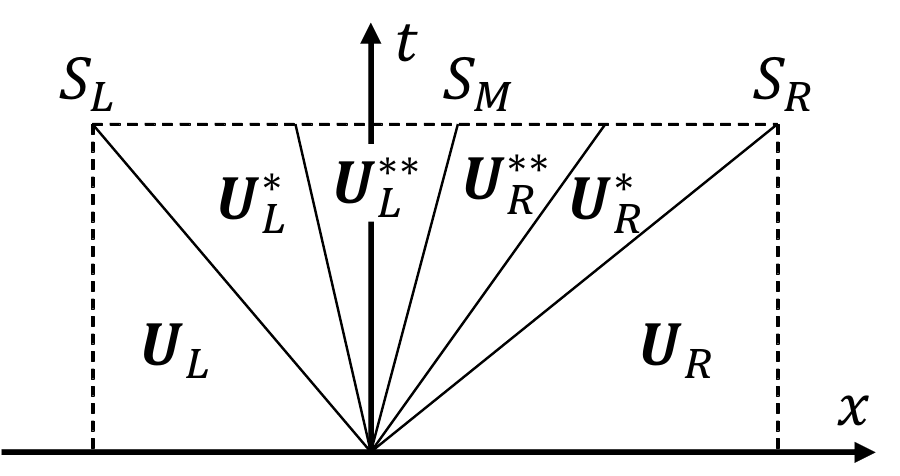}
\caption{Schematic picture of the Riemann fan with four intermediate states considered in the HLLD scheme.}
\label{fig:hlld_fig} 
\end{figure}

The HLLD scheme algebraically solves the MHD Riemann problem at a cell interface for the left- and right-side variables $\vect{U}_{L,R}$ as an initial state by allowing five eigenmodes in the Riemann fan (two fast modes, two {\Alfven} modes, and one entropy mode, shown in Figure \ref{fig:hlld_fig}).
The HLLD solution of density and tangential components in the outer sides in the Riemann fan (bounded by the fast and {\Alfven} modes) is written as:
\begin{eqnarray}
\rho_{\alpha}^* &=& \rho_{\alpha} \frac{S_{\alpha}-u_{\alpha}}{S_{\alpha}-S_M},\label{eq:24}\\
\vect{u}_{t,\alpha}^* &=& \vect{u}_{t,\alpha} - B_x \frac{S_M-u_{\alpha}}{X_{\alpha}} \vect{B}_{t,\alpha},\label{eq:25}\\
\vect{B}_{t,\alpha}^* &=& \tilde{\vect{B}}_{t,\alpha}+B_x^2 \frac{S_M-u_{\alpha}}{X_{\alpha}\left(S_{\alpha}-S_M\right)} \vect{B}_{t,\alpha},\label{eq:26}\\
\tilde{\vect{B}}_{t,\alpha}&=&\vect{B}_{t,\alpha}\frac{S_{\alpha}-u_{\alpha}}{S_{\alpha}-S_M},\label{eq:28}\\
X_{\alpha} &=& \rho_{\alpha} \left(S_{\alpha} - u_{\alpha}\right)\left(S_{\alpha} - S_M\right) - B_x^2,\label{eq:29}\\
S_L &=& \min\left(0,\min(u_L,u_R)-\max(c_{f,L},c_{f,R})\right),\label{eq:33}\\
S_R &=& \max\left(0,\max(u_L,u_R)+\max(c_{f,L},c_{f,R})\right),\label{eq:34}
\end{eqnarray}
where $\alpha = L \; {\rm or} \; R$ and $S_{L,R}$ are the minimum and maximum signal speeds.
The speed of the middle wave $S_M$ is calculated from the mass flux obtained in Section \ref{sec:mass-flux} as follows:
\begin{eqnarray}
 S_M = \frac{\dot{m}}{\rho} = \left\{
\begin{array}{l}
 {\dot{m}}/{\rho_L} = u_L, \;\;\; {\rm if} \; S_L =0,\\
 {\dot{m}}/{\rho_L^*} = \left(\dot{m} S_L\right)/\left[\dot{m} + \rho_L (S_L - u_L)\right], \;\;\; {\rm if} \; S_L S_R \neq 0 \; {\rm and} \; \dot{m} > 0,\\
 {\dot{m}}/{\rho_R^*} = \left(\dot{m} S_R\right)/\left[\dot{m} + \rho_R (S_R - u_R)\right], \;\;\; {\rm if} \; S_L S_R \neq 0 \; {\rm and} \; \dot{m} \leq 0,\\
 {\dot{m}}/{\rho_R} = u_R,. \;\;\; {\rm if} \; S_R =0.\\
\end{array}
\right.\label{eq:35}
\end{eqnarray}
$S_M$ at subsonic range $(S_L S_R \neq 0)$ is not necessarily identical to the one used in the HLLD scheme, while it reduces to the one-side value $u_{\alpha}$ at supersonic range $(S_{\alpha}=0)$ and is consistent with the HLLD scheme.
Given the condition of the rotational and contact discontinuities, the HLLD solution of density and tangential components in the inner sides in the Riemann fan (bounded by the {\Alfven} and entropy modes) satisfies the following equalities:
\begin{eqnarray}
 \rho^{**}_L &=& \rho^{*}_L,\label{eq:36}\\
 \rho^{**}_R &=& \rho^{*}_R,\label{eq:37}\\
 \vect{u}^{**}_{t,L} &=& \vect{u}^{**}_{t,R} = \vect{u}^{**}_{t},\label{eq:38}\\
 \vect{B}^{**}_{t,L} &=& \vect{B}^{**}_{t,R} = \vect{B}^{**}_{t}.\label{eq:27}
\end{eqnarray}
Using above quantities, the HLLD flux of the tangential momentum in the inner sides in the Riemann fan is expressed as:
\begin{eqnarray}
 \rho_{\alpha}^{**} S_M \vect{u}^{**}_t - B_x \vect{B}_t^{**} &=& \dot{m}\left(d_L \vect{u}_{t,L} + d_R \vect{u}_{t,R} \right) \nonumber \\
&+& \dot{m}\left\{d_L \left(\vect{u}_{t,L}^* - \vect{u}_{t,L}\right) + d_R \left(\vect{u}_{t,R}^* - \vect{u}_{t,R}\right) \right\} \nonumber \\
&-& \frac{{\rm sgn}\left(B_x\right)}{\sqrt{\rho_L^{*}}+\sqrt{\rho_R^{*}}}\left[\sqrt{\rho_R^{*}} \left(|B_x|+\frac{\dot{m}}{\sqrt{\rho_R^*}}\right)\vect{B}_{t,L}^* + \sqrt{\rho_L^{*}} \left(|B_x|-\frac{\dot{m}}{\sqrt{\rho_L^*}}\right)\vect{B}_{t,R}^*\right]\nonumber \\
&-& \frac{\sqrt{\rho_L^* \rho_R^*}}{\sqrt{\rho_L^{*}}+\sqrt{\rho_R^{*}}}\left[\left\{|B_x|-\left(\frac{d_L}{\sqrt{\rho_L^*}} + \frac{d_R}{ \sqrt{\rho_R^*}}\right)|\dot{m}|\right\}\left(\vect{u}_{t,R}^*-\vect{u}_{t,L}^*\right)\right].\label{eq:23}
\end{eqnarray}
where $d_{L,R}=[1\pm{\rm sgn}(\dot{m})]/2$ (Equation (\ref{eq:53})).
The four terms correspond to the advection term, the compression term in the presence of $B_x$, the right- and left-propagating {\Alfven} wave term, and the numerical diffusion term with a scale of {\Alfven} speed.
Evidently, the first term corresponds to the mass flux (first term of Equation (\ref{eq:11})).
Therefore, it is reasonable to define the magnetic tension flux $(B_x \vect{B}_t)_{\rm Present}$ that involves the remaining second to fourth terms of Equation (\ref{eq:23}). The numerical flux of the tangential momentum $\vect{F}^u$ is:
\begin{eqnarray}
\vect{F}^u &=& \dot{m}\left(d_L \vect{u}_{t,L} + d_R \vect{u}_{t,R} \right) - \left(B_x \vect{B}_t\right)_{\rm Present},\label{eq:55}\\
\left(B_x \vect{B}_t \right)_{\rm Present} &=& -\dot{m}\left\{d_L \left(\vect{u}_{t,L}^* - \vect{u}_{t,L}\right) + d_R \left(\vect{u}_{t,R}^* - \vect{u}_{t,R}\right) \right\}\nonumber \\
&& + \left({\cal A}_L^{u} \vect{B}_{t,L}^* + {\cal A}_R^{u} \vect{B}_{t,R}^*\right)
+ D^{u}\left(\vect{u}_{t,R}^*-\vect{u}_{t,L}^*\right),\label{eq:30}\\
 {\cal A}_{L,R}^{u} &=& {\rm sgn}(B_x) {\rm min}\left[ |B_x|, {\rm max}\left\{0,\frac{\sqrt{\rho_{R,L}^*}\left(|B_x| \pm \dot{m}/\sqrt{\rho_{R,L}^*}\right)}{\sqrt{\rho_L^*}+\sqrt{\rho_R^*}}\right\}\right],\label{eq:31}\\
D^{u} &=& {\rm max}\left[0,\frac{\sqrt{\rho_L^* \rho_R^*}}{\sqrt{\rho_L^{*}}+\sqrt{\rho_R^{*}}}\left\{|B_x|-\left(\frac{d_L}{\sqrt{\rho_L^*}} + \frac{d_R}{\sqrt{\rho_R^*}}\right)|\dot{m}|\right\} \right], \label{eq:32}
\end{eqnarray}
where we utilize minimum and maximum functions to extend the expression to the flux in the outer sides in the Riemann fan.
As would be shown later, the expression includes the flux in all the states of the HLLD solution.

Similarly, the HLLD flux of the tangential magnetic field in the inner sides in the Riemann fan is expressed as:
\begin{eqnarray}
  \vect{B}_t^{**} S_M - B_x \vect{u}_t^{**} &=& S_M\left(d_L \tilde{\vect{B}}_{t,L} + d_R \tilde{\vect{B}}_{t,R}\right)\nonumber \\
&+& S_M\left\{d_L \left(\vect{B}_{t,L}^*-\tilde{\vect{B}}_{t,L}\right) + d_R \left(\vect{B}_{t,R}^*-\tilde{\vect{B}}_{t,R}\right)\right\}\nonumber \\
&-& \frac{{\rm sgn}\left(B_x\right)}{\sqrt{\rho_L^{*}}+\sqrt{\rho_R^{*}}}\left[\sqrt{\rho_L^{*}} \left(|B_x|+\sqrt{\rho_R^{*}}S_M\right)\vect{u}_{t,L}^* + \sqrt{\rho_R^{*}} \left(|B_x|-\sqrt{\rho_L^{*}}S_M\right)\vect{u}_{t,R}^*\right]\nonumber \\
&-& \frac{1}{\sqrt{\rho_L^{*}}+\sqrt{\rho_R^{*}}}\left[\left\{|B_x|-\left(d_L \sqrt{\rho_L^*} + d_R \sqrt{\rho_R^*}\right)|S_M|\right\}\left(\vect{B}_{t,R}^*-\vect{B}_{t,L}^*\right)\right].\label{eq:39}
\end{eqnarray}
Each term has the same meaning as presented in Equation (\ref{eq:23}); using the relation $S_M \tilde{\vect{B}}_{t,\alpha} = \dot{m} (\vect{B}_{t,\alpha}/\rho_{\alpha})$, the first term corresponds to the mass flux (first term of Equation (\ref{eq:11})).
Therefore, we define the magnetic tension flux $(B_x \vect{u}_t)_{\rm Present}$ to involve the second to fourth terms of Equation (\ref{eq:39}).
The numerical flux of the tangential magnetic field $\vect{F}^B$ is given by:
\begin{eqnarray}
\vect{F}^B &=&  S_M\left(d_L \tilde{\vect{B}}_{t,L} + d_R \tilde{\vect{B}}_{t,R}\right) - \left(B_x \vect{u}_t\right)_{\rm Present},\label{eq:56}\\
 \left(B_x \vect{u}_t\right)_{\rm Present} &=& - S_M\left\{d_L \left(\vect{B}_{t,L}^*-\tilde{\vect{B}}_{t,L}\right) + d_R \left(\vect{B}_{t,R}^*-\tilde{\vect{B}}_{t,R}\right)\right\}\nonumber \\
&& + \left({\cal A}_L^{B} \vect{u}_{t,L}^* + {\cal A}_R^{B} \vect{u}_{t,R}^*\right) + D^{B} \left(\vect{B}_{t,R}^*-\vect{B}_{t,L}^*\right),\label{eq:40}\\
 {\cal A}_{L,R}^{B} &=& {\rm sgn}(B_x) {\rm min}\left[ |B_x|, {\rm max}\left\{0,\frac{\sqrt{\rho_{L,R}^*}\left(|B_x| \pm \sqrt{\rho_{R,L}^*} S_M\right)}{\sqrt{\rho_L^*}+\sqrt{\rho_R^*}}\right\}\right],\label{eq:41}\\
D^{B} &=& \frac{D^u}{\sqrt{\rho_L^* \rho_R^*}}.\label{eq:42}
\end{eqnarray}
Similar to Equations (\ref{eq:55})-(\ref{eq:32}), the expression is applicable to the flux in the outer sides, as well as the inner sides in the Riemann fan.

It is straightforward to prove that Equations (\ref{eq:30}) and (\ref{eq:40}) are consistent with the HLLD solution for $\forall S_M$.
Considering the flux in the outer side in the Riemann fan, i.e., a super-{\Alfvenic} case $(S_M = \dot{m}/\rho_L^* > |B_x|/\sqrt{\rho_L^*})$.
Thus, the numerical flux of the tangential momentum in Equation (\ref{eq:55}) becomes:
\begin{eqnarray}
\dot{m}\vect{u}_{t,L}-\left(B_x \vect{B}_t\right)_{\rm Present} = \dot{m}\vect{u}^*_{t,L} - B_x \vect{B}_{t,L}^*,\label{eq:43}
\end{eqnarray}
which is identical to the HLLD flux (except for the expression of $\dot{m}$).
Furthermore, it is rewritten as:
\begin{eqnarray}
\dot{m}\vect{u}_{t,L}-\left(B_x \vect{B}_t\right)_{\rm Present} = \dot{m} \vect{u}_{t,L}  - B_x \left[1 + \frac{\rho_L^* S_L(S_M-u_L)}{X_L}\right]\vect{B}_{t,L}.\label{eq:44}
\end{eqnarray}
Since $\dot{m} = \rho_L u_L$ as $S_L \rightarrow 0 \; (u_L>c_f,u_R>c_f)$ (Equations (\ref{eq:70})),
Equation (\ref{eq:44}) reduces to the one-side value $(\rho_L u_L \vect{u}_{t,L} - B_x \vect{B}_{t,L})$ for a super-fast-magnetosonic case.
The same holds for the tangential magnetic field.

Finally, we derive the magnetic tension flux for the total energy from the HLLD solutions as:
\begin{eqnarray}
\left(B_x \vect{u}_{t} \cdot \vect{B}_{t} \right)_{\rm Present} &=& \sgn(B_x)\left\{\frac{|B_x|}{S_\alpha-S_M} \left[{S_\alpha (\vect{u}^*_{t,\alpha} \cdot \vect{B}^*_{t,\alpha})} - {S_M (\vect{u}_{t,\alpha} \cdot \vect{B}_{t,\alpha})}\right]\right. \nonumber \\
 && \left. + \max(|B_x| - \sqrt{\rho^*_\alpha} |S_M|,0) \left(\vect{u}^{**}_{t} \cdot \vect{B}^{**}_{t} - \vect{u}^*_{t,\alpha} \cdot \vect{B}^*_{t,\alpha}\right)\right\},\label{eq:46}\\
\vect{u}_t^{**} \cdot \vect{B}_t^{**} &=&  \frac{\left(S_M \vect{F}^u + B_x \vect{F}^B \right) \cdot \left(B_x \vect{F}^u + \dot{m} \vect{F}^B \right)}{\left(\dot{m} S_M - B_x^2\right)^2},\label{eq:54}
\end{eqnarray}
where $\alpha= L \; ({\rm if} \; S_M > 0)\; {\rm or} \; R \; ({\rm otherwise})$.
Although it is nonidentical to Equation (\ref{eq:46}), one can alternatively use a scalar product of Equations (\ref{eq:30}) and (\ref{eq:40}) to reduce a calculation cost as:
\begin{eqnarray}
 \left(B_x \vect{u}_{t} \cdot \vect{B}_{t} \right)_{\rm Present} = \frac{\left(B_x \vect{u}_t\right)_{\rm Present} \cdot \left(B_x \vect{B}_t\right)_{\rm Present}}{B_x}.\label{eq:45}
\end{eqnarray}

\subsection{Pressure Flux}\label{sec:pressure-flux}
Most of AUSM-family schemes adopt the pressure flux in Equation (\ref{eq:8}), which is rewritten for low Mach number flows as:
\begin{eqnarray}
\hat{P}_{i+1/2} &=& \bar{P}_{i+1/2} - \frac{{\cal P}_+ - {\cal P}_-}{2}\Delta P_{i+1/2} + \left({\cal P}_+ + {\cal P}_- - 1\right) \bar{P}_{i+1/2}\label{eq:19}\\
&\simeq& \bar{P}_{i+1/2} - \frac{15}{16}\bar{M}_{i+1/2} \Delta P_{i+1/2} - \frac{15}{16} \bar{P}_{i+1/2} \Delta M_{i+1/2},\label{eq:20}
\end{eqnarray}
where $\bar{X}_{i+1/2} = (X_{i+1/2,L} + X_{i+1/2,R})/2$.
\cite{2011AIAAJ..49.1693S} argued that the last term of Equation (\ref{eq:20}) acts as diffusion with a scale of sound speed, $P\Delta M = \rho a \Delta u /\gamma$, and it is too large for low Mach number flows.
Then, they proposed simple low-dissipation AUSM-family schemes to correct the pressure flux, in which a function $f(M)  = \min(|M|(2-|M|),1)$ \citep[SLAU;][]{2011AIAAJ..49.1693S} or $f(M) = |M|$ \citep[SLAU2;][]{2013JCoPh.245...62K} is multiplied to the last term of Equation (\ref{eq:19}) to scale the diffusion with the advection speed rather than with the sound speed (note that they adopt a multidimensional Mach number $|M|=|\vect{u}|/a$ rather than a normal Mach number $|u|/a$ based on numerical investigations).
 \cite{2006JCoPh.214..137L} extended the AUSM$^+$-up scheme to solve low speed flows by correcting the pressure flux in a similar manner based on the asymptotic analysis for low Mach numbers.
Expanding the pressure with respect to the Mach number and taking its low Mach number limit, he argued that the pressure flux in Equation (\ref{eq:19}), $\hat{P}=\bar{P}+O(M^1)$, should be corrected to match the asymptotic solution $P(x,t)=P^{(0)}(t)+M^2P^{(2)}(x,t)+O(M^3)$.
This is a strategy to achieve all-speed AUSM-family schemes for hydrodynamic simulations.

\cite{2019ShWav..29..611K} applied the pressure flux of the SLAU2 scheme to MHD simulations by replacing the gas pressure with the total pressure $P_t=P+|\vect{B}_t|^2/2$ and the Mach number with the fast magnetosonic Mach number, and setting $f(M)=|\vect{u}|/c_f$ as follows:
\begin{eqnarray}
\hat{P}_{t,i+1/2}^{\rm SLAU2} &=& \bar{P}_{t,i+1/2} - \frac{{\cal P}_+ - {\cal P}_-}{2}\Delta P_{t,i+1/2} + \frac{|\vect{u}_{i+1/2}|}{c_{f,i+1/2}}\left({\cal P}_+ + {\cal P}_- - 1\right) \bar{P}_{t,i+1/2}.\label{eq:73}
\end{eqnarray}
The factor $|\vect{u}|/c_f$ decreases the interface pressure to improve the resolution of low Mach number flows, and increases it to stabilize high Mach number flows.
The approach is inspired by a preconditioning technique, which alters the propagation speed $u \pm a$ by $u \pm f(M)a \sim u \pm |u|$ to relax the stiffness of the hyperbolic system of equations (Appendix \ref{sec:prec-syst-euler}).
Although it works well for hydrodynamic simulations \citep{2013JCoPh.245...62K}, its application to MHD simulations is still questionable because of the presence of the {\Alfven} and slow modes, which should be slower than the fast mode.
Consequently, we correct the pressure flux as:
\begin{eqnarray}
\hat{P}_{t,i+1/2}^{\rm Present} &=& \bar{P}_{t,i+1/2} - \frac{{\cal P}_+(M_{i+1/2,L}) - {\cal P}_-(M_{i+1/2,R})}{2}\Delta P_{t,i+1/2} \nonumber \\
 &+& \frac{c_{u,i+1/2}}{c_{f,i+1/2}} \left({\cal P}_+(M_{i+1/2,L}) + {\cal P}_-(M_{i+1/2,R}) - 1\right) \bar{P}_{t,i+1/2}\nonumber \\
 &-& \frac{1}{2}{\cal P}_+(M_{i+1/2,L}){\cal P}_-(M_{i+1/2,R})\bar{\rho}_{i+1/2}c_{u,i+1/2}\Delta u_{i+1/2},\;\;\;M_{i+1/2,\alpha}=\frac{u_{i+1/2,\alpha}}{c_{f,i+1/2}},\nonumber \\
  \label{eq:21}
\end{eqnarray}
where,
\begin{eqnarray}
 c_{u,i+1/2}=\max(c_{u,i+1/2,L},c_{u,i+1/2,R}), \; c_u^2 = \frac{1}{2}\left[\left(c^2_a + |\vect{u}|^2\right) + \sqrt{\left(c^2_a + |\vect{u}|^2\right)^2-4|\vect{u}|^2 c^2_{ax}}\right], \nonumber \\
 \label{eq:22}
\end{eqnarray}
is the modified fast magnetosonic speed that remains faster than the {\Alfven} and slow magnetosonic speeds.
The velocity difference term (fourth term of Equation (\ref{eq:21})) is taken from the AUSM$^+$-up scheme to enhance the stability for low Mach number flows.
The pressure flux can be effectively decreased in high beta and low speed plasma $(c_u \ll c_f)$, but not in low beta plasma.
The scheme is expected to properly scale down to $|\vect{u}| = c_a$ because the third and fourth dissipation terms are approximated by $-{\rm max}(|\vect{u}|,c_a) \rho \Delta u$.
Note that the above-stated pressure flux is essentially the correction of the amount of numerical dissipation, which as opposed to the preconditioning technique, it does not alter the basic equations themselves.

For comparison, we describe the pressure flux of the familiar hydrodynamic shock-capturing schemes.
The Roe scheme uses:
\begin{eqnarray}
P_{1/2}^{\rm Roe} &=& \left(\rho u^2 + P\right)_{1/2}-u_{1/2}\left(\rho u\right)_{1/2}\nonumber \\
&=& \bar{P}-\left(\frac{|u+a|-|u-a|}{2a}\right)_{1/2}\Delta P - \left(\frac{|u+a|+|u-a|}{2}\right)_{1/2}\rho_{1/2} \Delta u \nonumber \\
&=& \bar{P}-M_{1/2}\Delta P - \rho_{1/2} a_{1/2} \Delta u, \;\;\; {\rm for} \; M<1,\label{eq:75}
\end{eqnarray}
and the HLLC scheme uses:
\begin{eqnarray}
P_{1/2}^{\rm HLLC} \simeq \frac{\rho_{L}P_{L} + \rho_{R}P_{R}}{\rho_{L} + \rho_{R}} - \frac{\rho_{L} \rho_{R}}{\rho_{L} + \rho_{R}} a_{1/2} \Delta u,\;\;\; {\rm for} \; M<1.\label{eq:67}
\end{eqnarray}
Both schemes have a velocity difference term with a scale of sound speed similar to Equation (\ref{eq:20}), and it is too large for low Mach number flows.
The same holds for MHD shock-capturing schemes.

\subsection{Interpolation and Extension to Multidimension}\label{sec:interp-extens-mult}
Left and right states at a cell interface are interpolated by higher-order polynomials to obtain a better accuracy of the solution.
In compressible fluid simulations, nonlinear interpolation methods are usually employed to avoid unphysical oscillation around discontinuous regions.
For example, the second-order MUSCL scheme developed by \cite{1979JCoPh..32..101V} has been widely adopted for practical MHD simulations.

There are several candidate variables for interpolation.
It is more appropriate for upwind schemes (including AUSM-family schemes) to interpolate characteristic variables rather than conservative or primitive variables \citep{2002JCoPh.183..187Q,2009SIAMR..51...82S,2019PASJ...71...83M}.
However, we find that the characteristic decomposition combined with the present scheme using the pressure flux correction (Equation (\ref{eq:21})) causes unphysical oscillation in very low Mach number flows, irrespective of the presence of the magnetic field.
This is probably due to the inconsistency that the scheme approaches a central scheme in low Mach number limits, while the characteristic decomposition is based on the hyperbolicity of the system and is appropriate for upwind schemes (note that the characteristic decomposition works well if the pressure flux is not corrected).
Since the characteristic decomposition is quite useful for supersonic flows to suppress numerical oscillation, while the pressure flux is corrected only for low Mach number flows, we use the characteristic decomposition provided that a flow speed exceeds the sound speed $|\vect{u}|/a > 1$ in a stencil $(i-1,i,i+1)$ (see Section 2.3 in \cite{2019ApJS..242...14M} for details of the characteristic decomposition).
Otherwise, we use the following approximate characteristic variables $d\vect{W}$ for the interpolation:
\begin{eqnarray}
\begin{pmatrix}
 dW_1\\
dW_2\\
dW_3\\
dW_4\\
dW_5\\
dW_6\\
dW_7
\end{pmatrix}
=
\begin{bmatrix}
 0& 1& 0& 0& 0& 0& 0\\
c_f^2 & 0 & 0 & 0 & -B_y & -B_z & -1\\
0&0&0&0&B_y&B_z&1\\
0&0&\sqrt{\rho}&0&1&0&0\\
0&0&\sqrt{\rho}&0&-1&0&0\\
0&0&0&\sqrt{\rho}&0&1&0\\
0&0&0&\sqrt{\rho}&0&-1&0\\
\end{bmatrix}
\begin{pmatrix}
 d\rho\\
du\\
dv\\
dw\\
dB_y\\
dB_z\\
dP
\end{pmatrix}
,\label{eq:57}\\
\begin{pmatrix}
 d\rho\\
du\\
dv\\
dw\\
dB_y\\
dB_z\\
dP
\end{pmatrix}
=
\begin{bmatrix}
 0& \frac{1}{c_f^2}& \frac{1}{c_f^2}& 0& 0& 0& 0\\
1 & 0 & 0 & 0 & 0 & 0 & 0\\
0&0&0&\frac{1}{2\sqrt{\rho}}&\frac{1}{2\sqrt{\rho}}&0&0\\
0&0&0&0&0&\frac{1}{2\sqrt{\rho}}&\frac{1}{2\sqrt{\rho}}\\
0&0&0&\frac{1}{2}&-\frac{1}{2}&0&0\\
0&0&0&0&0&\frac{1}{2}&-\frac{1}{2}\\
0&0&1&-\frac{B_y}{2}&\frac{B_y}{2}&-\frac{B_z}{2}&\frac{B_z}{2}
\end{bmatrix}
\begin{pmatrix}
 dW_1\\
dW_2\\
dW_3\\
dW_4\\
dW_5\\
dW_6\\
dW_7
\end{pmatrix}
,\label{eq:59}
\end{eqnarray}
where $dW_3 = dP_t$, $dW_2$ and $dW_{4-7}$ approximate the entropy and {\Alfven} modes, respectively.

The present scheme can be extended to multidimension through a dimension-by-dimension reconstruction.
However, it is widely recognized that multidimensional MHD simulations should exercise special care with the divergence-free condition for the magnetic field to obtain reliable solutions.
To preserve the divergence-free condition within a machine precision, we adopt the central upwind constrained transport method proposed by \cite{2019ApJS..242...14M}, which is built to be consistent with the base one-dimensional scheme and attain a designed high order of accuracy in multidimension.

\subsection{Summary of Flux Function}\label{sec:summ-num-flux}
Here, we summarize the numerical flux function of the present AUSM-family scheme.
The scheme splits the inviscid flux in MHD equations into the advection, pressure, and magnetic tension parts at the cell interface (Equation (\ref{eq:11})), and then individually evaluates the mass, pressure, and magnetic tension fluxes.
The mass flux is presented in Equations (\ref{eq:70})-(\ref{eq:17}), and it includes the shock detection to deal with the numerical instability (Equations (\ref{eq:51})-(\ref{eq:50})).
The magnetic tension flux in Equations (\ref{eq:30})-(\ref{eq:32}), (\ref{eq:40})-(\ref{eq:42}), and (\ref{eq:46})-(\ref{eq:54}) is built to be consistent with the HLLD scheme to preserve the rotational discontinuity.
The pressure flux in Equations (\ref{eq:21})-(\ref{eq:22}) scales properly for low Mach number flows to improve the resolution.
Hereafter, this scheme is termed multistate low-dissipation AUSM (MLAU).

The MLAU scheme inherits the property of the exact preservation of the stationary MHD discontinuities from the baseline AUSM$^+$-up and HLLD schemes as follows:
\begin{itemize}
 \item Contact discontinuity. Let, $[\rho,u,v,w,B_y,B_z,P]=[\rho_L,0,v_0,w_0,B_{y0},B_{z0},P_0]\; (x \leq 0)$, $[\rho_R,0,v_0,w_0,B_{y0},B_{z0},P_0]\; (x > 0)$, and $B_x \neq 0$. 
The resultant numerical fluxes, $\dot{m}=0,\hat{P}_{t,1/2}=P_0+(B_{y0}^2+B_{z0}^2)/2,\hat{\vect{T}}_{1/2}=B_x(0,B_x/2,B_{y0},B_{z0},v_0,w_0,v_0B_{y0}+w_0B_{z0})^T$, are exact.
 \item Tangential discontinuity. Let, $[\rho,u,v,w,P_t]=[\rho_L,0,v_L,w_L,P_{t0}]\; (x \leq 0)$, $[\rho_R,0,v_R,w_R,P_{t0}]\; (x > 0)$, and $B_x = 0$. 
The resultant numerical fluxes, $\dot{m}=0,\hat{P}_{t,1/2}=P_{t0},\hat{\vect{T}}_{1/2}=\vect{0}$, are exact.
 \item Rotational discontinuity. Let, $[\rho,u,\vect{u}_t,\vect{B}_t,P]=[\rho_0,-B_x/\sqrt{\rho_0},\vect{u}_{t,L},\vect{B}_{t,L},P_0]\; (x \leq 0)$, $[\rho_0,-B_x/\sqrt{\rho_0},\vect{u}_{t,R},\vect{B}_{t,R},P_0]\; (x > 0)$, and $\Delta \vect{u}_t = -\Delta \vect{B}_t/\sqrt{\rho_0}, |\vect{B}_{t,L}|^2=|\vect{B}_{t,R}|^2=B_t^2,B_x > 0$.
The resultant numerical fluxes are $\dot{m}=-\sqrt{\rho_0}B_x,\hat{P}_{t,1/2}=P_{0}+B_t^2/2,\hat{\vect{T}}_{1/2}=B_x(0,B_x/2,\vect{B}_{t,R},\vect{u}_{t,R},\vect{u}_{t,R} \cdot \vect{B}_{t,R})^T$, and then,
\begin{eqnarray}
 \hat{\vect{F}}_{1/2} &=& 
\begin{bmatrix}
 -\sqrt{\rho_0}B_x\\
 P_0 + (B_x^2+B_t^2)/2\\
-\sqrt{\rho_0}B_x\left(\vect{u}_{t,R}+\vect{B}_{t,R}/\sqrt{\rho_0}\right)\\
             -B_x\left(\vect{u}_{t,R}+\vect{B}_{t,R}/\sqrt{\rho_0}\right)\\
-\sqrt{\rho_0}B_x\left\{\frac{\gamma P_0}{(\gamma-1)\rho_0} + \frac{(B_x^2+B_t^2)}{2} + \frac{\left|\vect{u}_{t,R}+\vect{B}_{t,R}/\sqrt{\rho_0}\right|^2}{2}\right\}
\end{bmatrix}
\nonumber \\ 
&=&
\begin{bmatrix}
 -\sqrt{\rho_0}B_x\\
 P_0 + (B_x^2+B_t^2)/2\\
-\sqrt{\rho_0}B_x\left(\vect{u}_{t,L}+\vect{B}_{t,L}/\sqrt{\rho_0}\right)\\
             -B_x\left(\vect{u}_{t,L}+\vect{B}_{t,L}/\sqrt{\rho_0}\right)\\
-\sqrt{\rho_0}B_x\left\{\frac{\gamma P_0}{(\gamma-1)\rho_0} + \frac{(B_x^2+B_t^2)}{2} + \frac{\left|\vect{u}_{t,L}+\vect{B}_{t,L}/\sqrt{\rho_0}\right|^2}{2}\right\}
\end{bmatrix}
\label{eq:72}
\end{eqnarray}
are exact.
\end{itemize}

\section{Numerical Tests}\label{sec:numerical-tests}
This section presents the numerical simulation results of MHD test problems to assess the capability of the MLAU scheme through comparison with the standard Roe and HLLD schemes.
In the following one-dimensional shock tube tests (\S \ref{sec:one-dimens-shock}), the order of accuracy of the scheme is first in space and second in time to focus on the essence of the flux function.
In the remaining tests (\S \ref{sec:line-wave-prop}-\ref{sec:richtmy-meshk-inst}), we employ the MUSCL interpolation with a minmod limiter and the third-order strong stability preserving Runge-Kutta method \citep{1988JCoPh..77..439S,1998MaCom..67...73G}.
We solve ideal MHD equations with a specific heat ratio $\gamma=5/3$ and a CFL number of 0.4, unless otherwise stated.

\subsection{One-dimensional Shock Tube Problems}\label{sec:one-dimens-shock}
The shock tube problem is a standard test for compressible fluid simulations to assess the capability of capturing discontinuities and rarefaction waves with satisfactory accuracy and robustness.
We conduct three one-dimensional problems adopted by \cite{2005JCoPh.208..315M}.
The computational domain of $-0.5<x<0.5$ is resolved by 800 cells, and 
the initial state is separated at $x=0$.
The open boundary condition is imposed for characteristic variables.

\begin{figure}[htbp]
\centering
\includegraphics[clip,angle=0,scale=0.35]{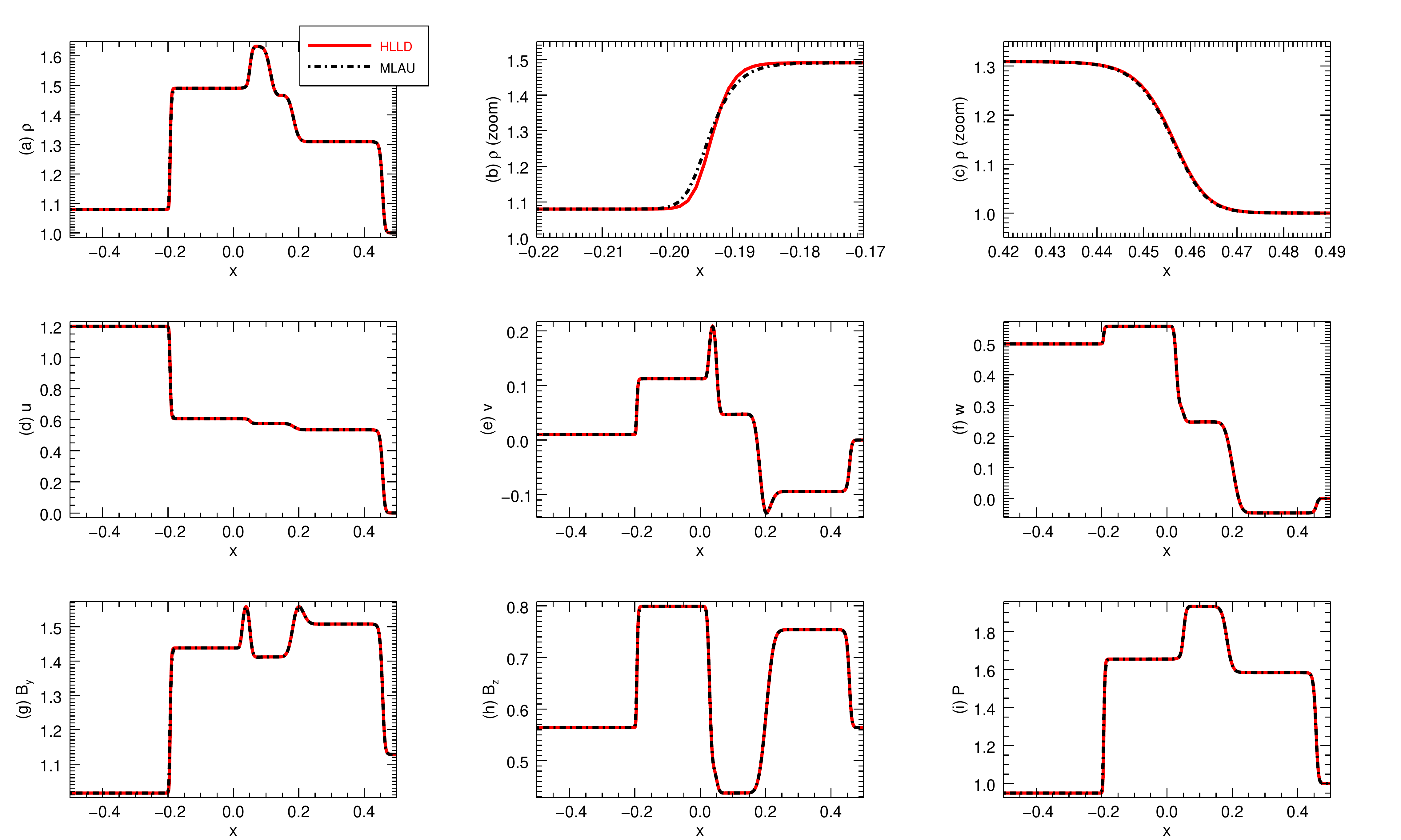}
\caption{One-dimensional shock tube problem of \cite{1994JCoPh.111..354D}. The solid red and dashed black lines are obtained with the HLLD and MLAU schemes. Panels (b) and (c) are magnifications of the density profile around fast shocks at $x=-0.2$ and $x=0.45$.}
\label{fig:dwshock1d} 
\end{figure}

The first problem is one of the familiar shock tube problem presented by \cite{1994JCoPh.111..354D}.
The initial left and right states are $[\rho,u,v,w,B_{y},B_{z},P] = [1.08,1.2,0.01,0.5,3.6/\sqrt{4 \pi},2/\sqrt{4 \pi},0.95]$ and $[1,0,0,0,4/\sqrt{4 \pi},2/\sqrt{4 \pi},1]$ with $B_x=2/\sqrt{4 \pi}$.
Figure \ref{fig:dwshock1d} shows the primitive variables at $t=0.2$ obtained with the HLLD (red) and MLAU (black) schemes.
The solution with the Roe scheme is almost identical to the HLLD scheme, and thus, it is not shown here.
The solution to this problem includes all MHD discontinuities, fast shocks $(x=-0.2,0.45)$, rotational discontinuities $(x=0,0.2)$, slow shocks $(x=0.05,0.15)$, and a contact discontinuity $(x=0.1)$.
The MLAU scheme captures all discontinuities well and its solution is very similar to the HLLD scheme.
A slight difference is found in the density profile around the fast shock.
The solution with the MLAU scheme is more diffusive than the HLLD scheme at the left-hand fast shock in panel (b), while they are comparable at the right-hand fast shock in panel (c).
We confirm that the thickness of the shock in the MLAU scheme is sensitive to the amount of the pressure difference term in the mass flux (Equation (\ref{eq:16})) and the velocity difference term in the pressure flux (Equation (\ref{eq:21})), and decreasing their coefficients steepens the shock.

\begin{figure}[htbp]
\centering
\includegraphics[clip,angle=0,scale=0.35]{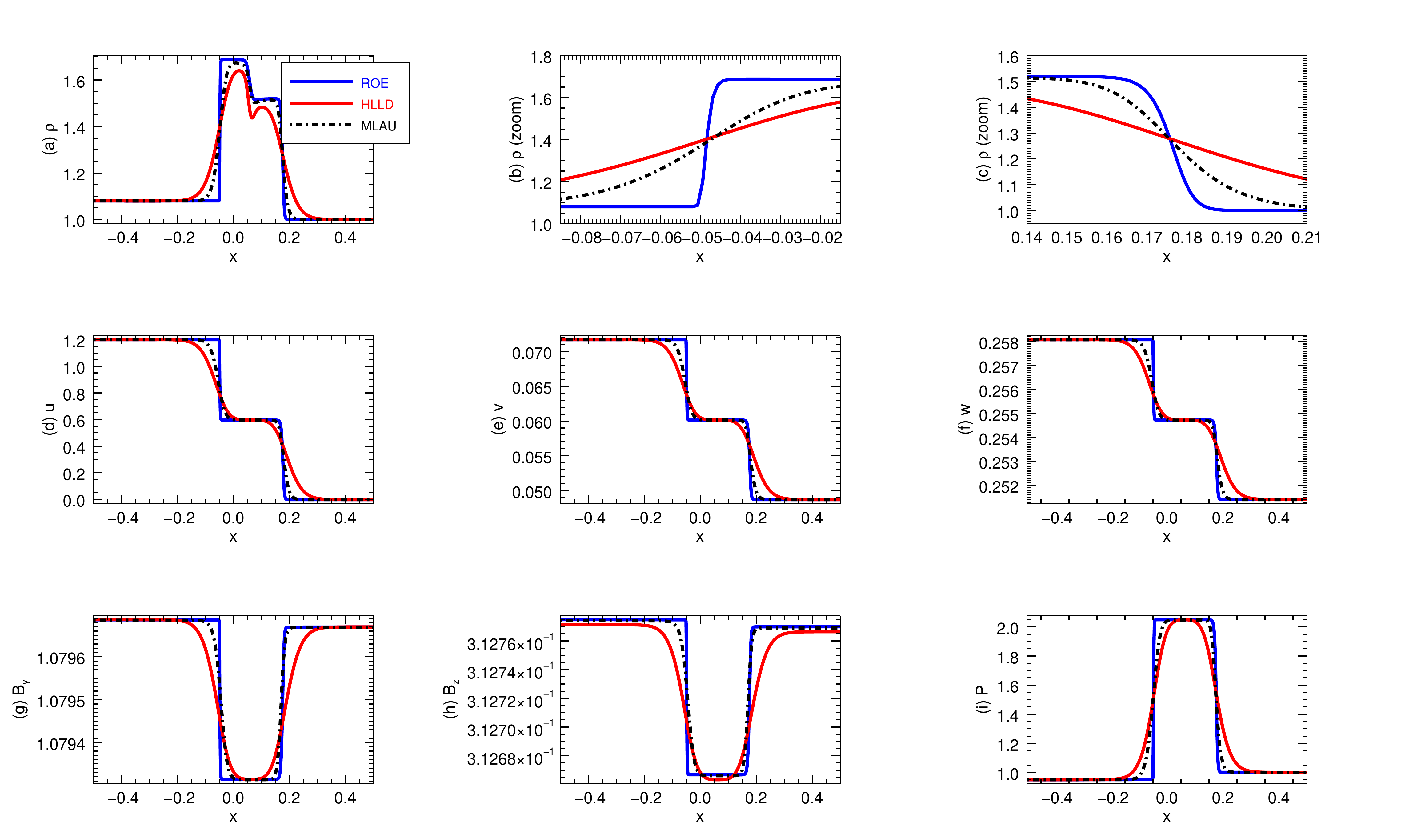}
\caption{Same as Figure \ref{fig:dwshock1d}, but $B_x$ is a hundredfold to focus on the slow mode shock.}
\label{fig:dwshock1d_100} 
\end{figure}

One of the main differences among the Roe, HLLD, and MLAU schemes is the treatment of the slow mode wave. The Roe scheme satisfies the jump condition for all eigenmodes, while the HLLD scheme appropriately omits information on the slow mode, and the MLAU scheme is neglectful about it.
To check their effects on the resolution of the slow mode, we conduct the same shock tube problem with an extremely strong $B_x=200/\sqrt{4 \pi}$.
Figure \ref{fig:dwshock1d_100} compares the solution of the three schemes.
As expected, the HLLD scheme (red) smears slow mode shocks at $x=-0.05,0.18$, owing to excessive numerical diffusion with a scale of {\Alfven} speed, while the Roe scheme (blue) correctly resolves them.
The MLAU scheme (black) also fails to resolve the slow mode shocks, but the thickness is steeper than in the HLLD scheme.
We confirm that the thickness of these slow mode shocks is sensitive to the amount of the velocity difference term in the pressure flux (Equation (\ref{eq:21})), which is approximated as $-\rho c_f \Delta u/8$ in low plasma beta $(2P/|\vect{B}|^2 \ll 1)$ and low Mach number limit. 
Meanwhile, the HLLD scheme also has a velocity difference term in the interface pressure (Equation (41) in \cite{2005JCoPh.208..315M}), and the term is approximately four times larger than the MLAU scheme in the low beta and low Mach number limit.

\begin{figure}[htbp]
\centering
\includegraphics[clip,angle=0,scale=0.35]{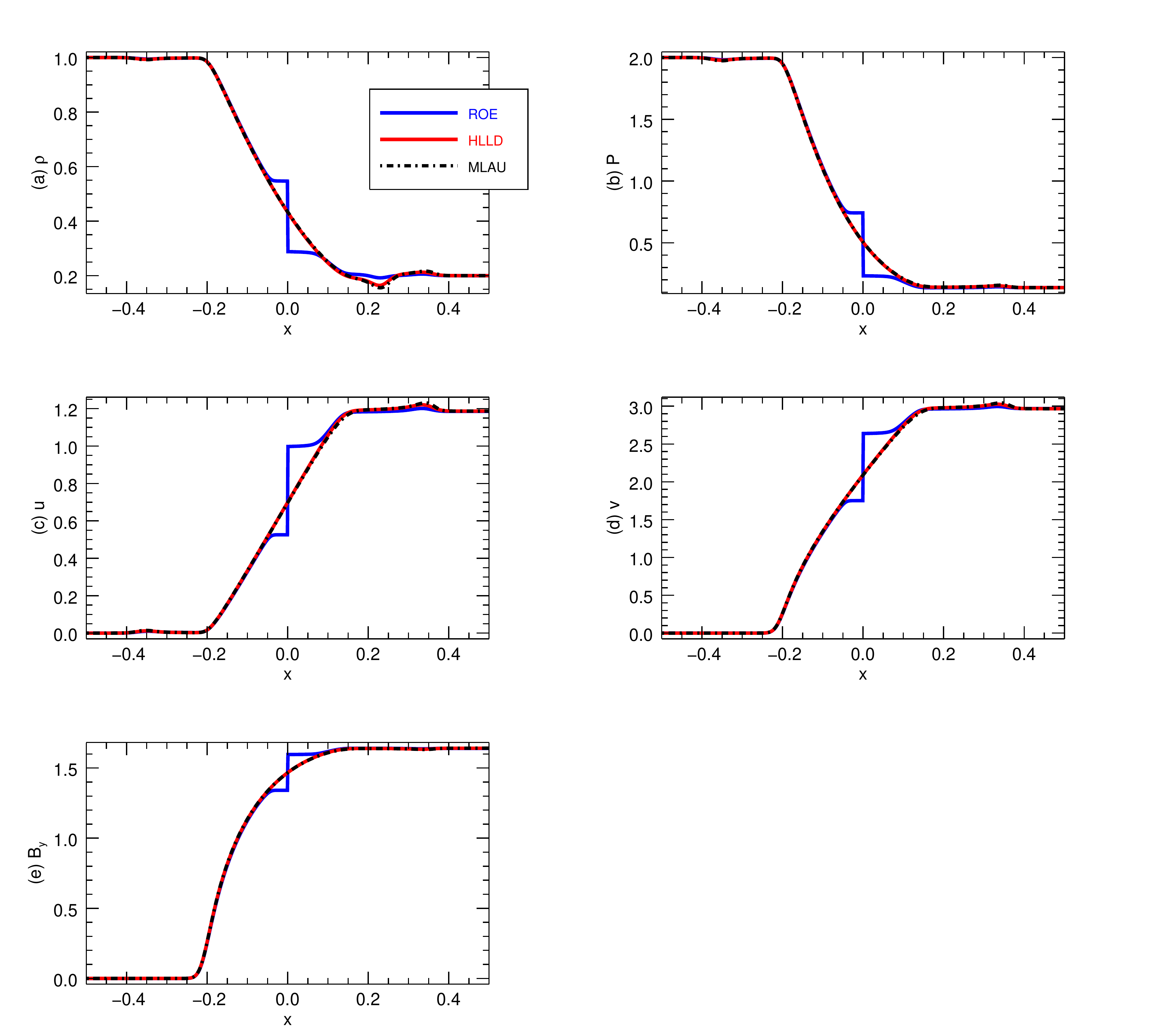}
\caption{One-dimensional switch-off slow rarefaction wave. The format is the same as Figure \ref{fig:dwshock1d}.}
\label{fig:ssrare} 
\end{figure}

The second problem is a switch-off slow rarefaction wave problem.
The initial left and right states are $[\rho,u,v,w,B_{y},B_{z},P] = [1,0,0,0,0,0,2]$ and $[0.2,1.186,2.967,0,1.6405,0,0.1368]$ with $B_x=1$.
\cite{2005JCoPh.208..315M} verified that the HLLD scheme resolves the slow rarefaction wave with satisfactory accuracy, while the Roe scheme provides an unphysical solution to this problem due to the violation of the entropy condition at $x=0$ and requires an additional entropy correction to cure the solution.
Figure \ref{fig:ssrare} compares the solution at $t=0.2$ among the three schemes.
The MLAU scheme does not violate the entropy condition and its solution is close to the HLLD scheme.

\begin{figure}[htbp]
\centering
\includegraphics[clip,angle=0,scale=0.35]{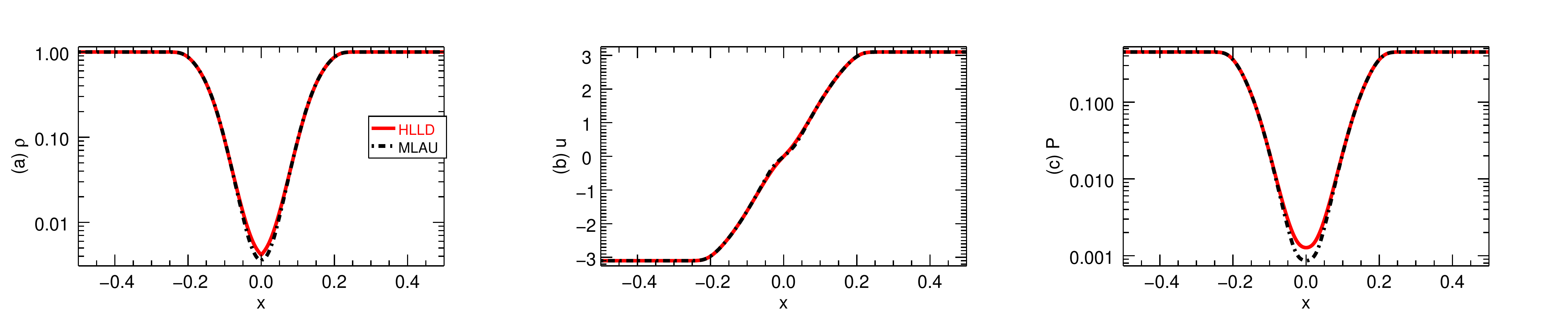}
\caption{One-dimensional super-fast expansion with a Mach number of 3.1. The format is the same as Figure \ref{fig:dwshock1d}.}
\label{fig:sfexpan} 
\end{figure}

The last problem is a super-fast expansion wave problem.
The initial left and right states are $[\rho,u,v,w,B_{y},B_{z},P] = [1,-u_0,0,0,0.5,0,0.45]$ and $[1,u_0,0,0,0.5,0,0.45]$ with $B_x=0$ and $u_0=3.1$ is equal to the fast magnetosonic Mach number.
The HLLD scheme can solve this problem because of the positivity preservation, while the Roe scheme fails due to the violation of positivity.
Figure \ref{fig:sfexpan} shows the result at $t=0.05$ obtained with the HLLD and MLAU schemes.
The MLAU scheme also succeeds in this problem, indicating a better stability than obtainable with the Roe scheme.
We note that the MLAU scheme inherits the positivity preservation of density from the baseline AUSM$^+$-up scheme, but it does not guarantee the positivity of pressure.
Therefore, the scheme would not necessarily be robust as much as the HLLD scheme from the viewpoint of the positivity preservation.

\subsection{Linear wave propagation}\label{sec:line-wave-prop}
We simulate the propagation of linear MHD waves in one-dimensional homologous medium with the MLAU scheme.
The initial condition is uniform, $[\rho,u,v,w,B_{y},B_{z},P] = [1,u_0,0,0,0,0,\beta B_x^2/2]$ with $B_x = 1$.
Then, a small $(1 \%)$ uniform random perturbation is added to $B_y$ and $P$.
We use the ambient flow velocity $u_0=(0,0.1c_f,1.5c_f)$ and the plasma beta $\beta=(10,0.1)$ (a total of six simulation runs).
The periodic computational domain of $0<x<L_x=1$ is resolved by 256 cells.

\begin{figure}[htbp]
\centering
\includegraphics[clip,angle=0,scale=0.35]{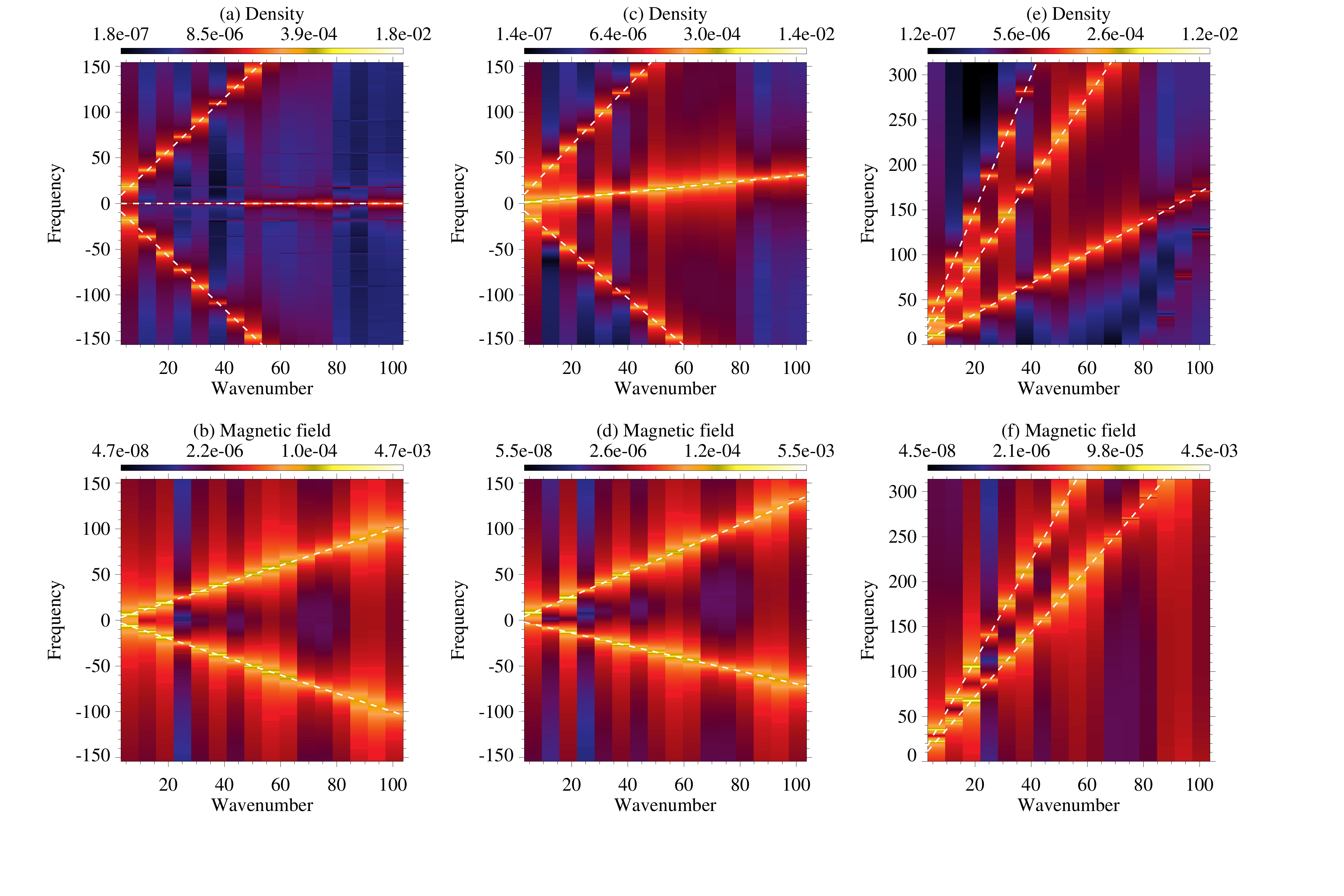}
\caption{Dispersion relation of MHD eigenmodes for $\beta=10$. Upper panels represent the Fourier amplitude of density, corresponding to the fast and the entropy modes. Lower panels are the Fourier amplitude of $B_y$, corresponding to the {\Alfven} mode. Panels (a)-(b) are for the stationary case $(u=0)$, (c)-(d) for the subsonic case $(u=0.1 c_f)$, and (e)-(f) for the supersonic case $(u=1.5 c_f)$, respectively. The dashed lines represent the theoretical dispersion relation.} 
\label{fig:hbeta_wave}
\end{figure}

\begin{figure}[htbp]
\centering
\includegraphics[clip,angle=0,scale=0.35]{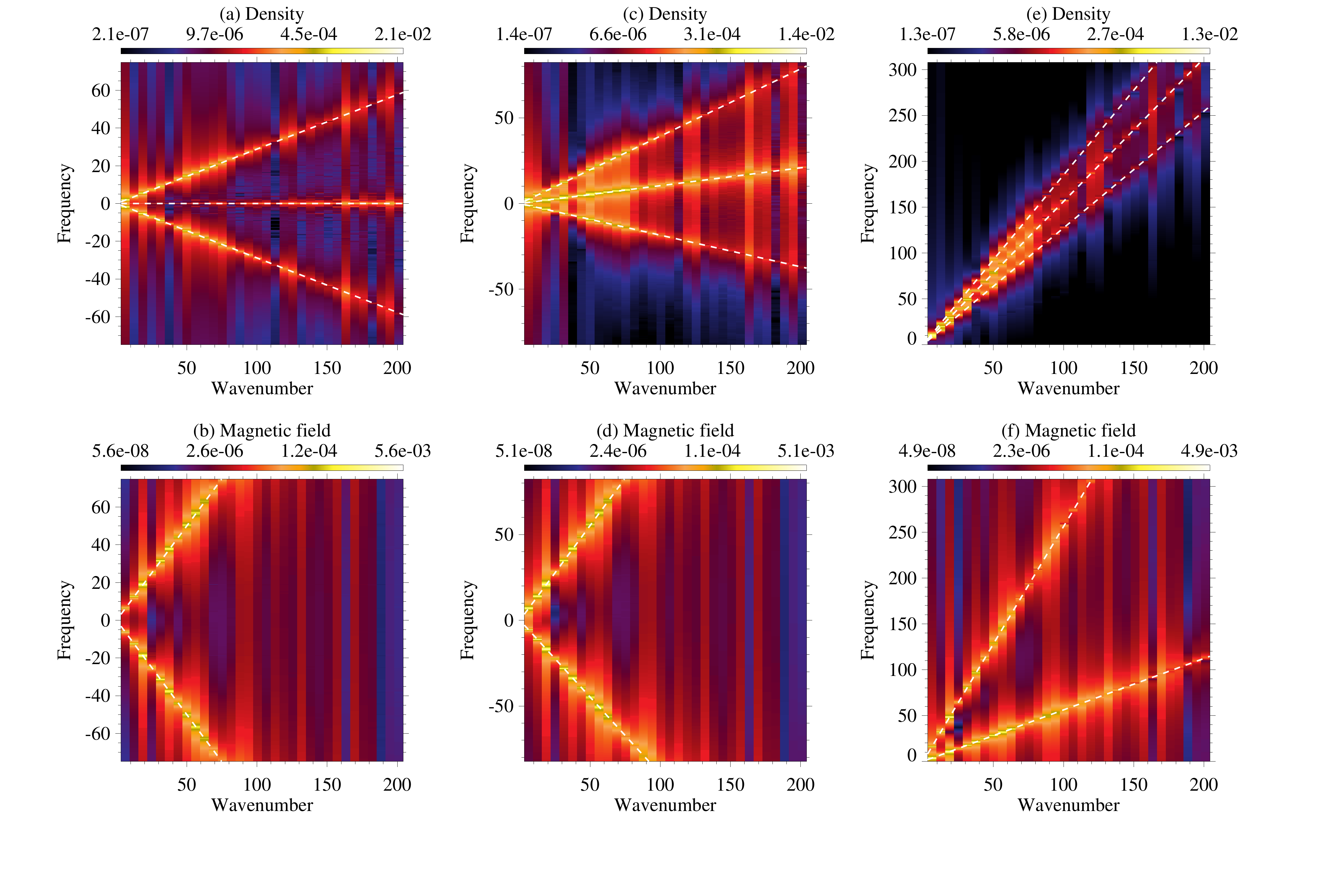}
\caption{Same as Figure \ref{fig:hbeta_wave}, but for $\beta=0.1$. Upper panels correspond to the slow and the entropy modes.} 
\label{fig:lbeta_wave}
\end{figure}

Figures \ref{fig:hbeta_wave} and \ref{fig:lbeta_wave} show the dispersion relation for high and low beta cases.
The upper panels correspond to the compressible mode (fast/slow modes for the high/low beta cases) and the entropy mode, while the lower panels correspond to the {\Alfven} mode.
The solution agrees well with the theoretical dispersion relation $\omega = \lambda k_x$ denoted by solid lines in each plasma frame (stationary in left, subsonic in middle, and supersonic in right panels), where $k_x,\omega$, and $\lambda$ are the wavenumber, frequency, and eigenvalues, respectively.
Note that the horizontal stripes seen in Figures \ref{fig:hbeta_wave}(a) and \ref{fig:lbeta_wave}(a) are pseudo entropy modes owing to a finite domain size, $\omega = uk_x \pm 2 \pi n c_{f,s}/L_x \; (n=1,2,\dots)$, irrespective of the choice of the flux function.
Any spurious growth of the wave is not observed until the simulation ends at $t=40$.
This problem confirms that the MLAU scheme does not alter the propagation speed of the eigenmodes, nor does it degrade their numerical stability.

\subsection{Orszag-Tang vortex}\label{sec:orszag-tang-vortex}
The Orszag-Tang vortex is a standard two-dimensional MHD test problem to verify the capability to capture multiple interactions of shock waves and vortices \citep{1979JFM....90..129O}.
The initial condition is $(\rho,u,v,w,B_x,B_y,B_z,P)=(\gamma^2,-\sin(y),\sin(x),0,-\sin(y),\sin(2x),0,\gamma)$.
The periodic computational domain of $0<x<2\pi$ and $0<y<2\pi$ is resolved by $N\times N$ cells, where $N=200$ as a fiducial run and $N=1600$ as a reference run.

\begin{figure}
\centering
\includegraphics[clip,angle=0,scale=0.25]{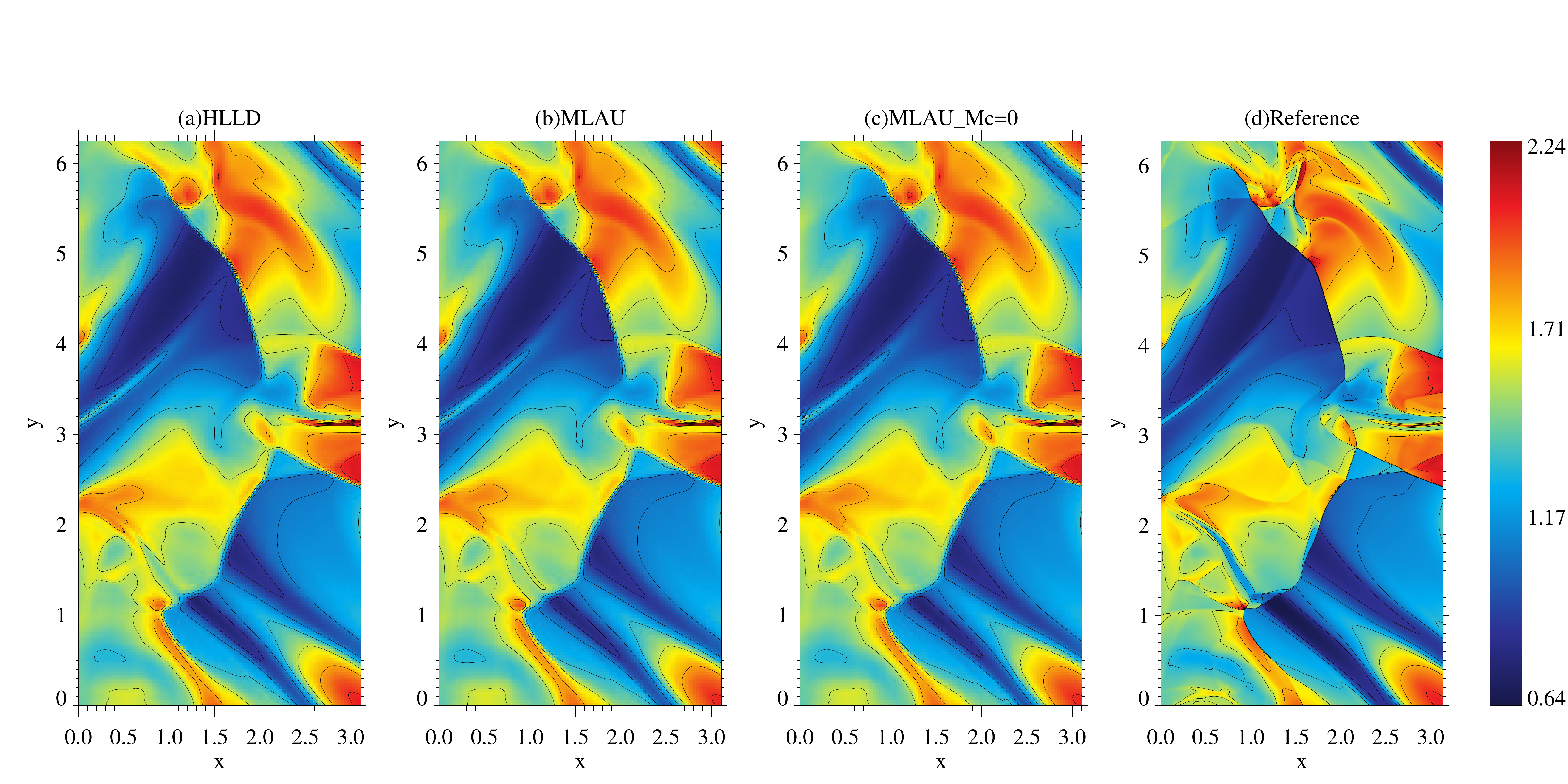}
\caption{Temperature profile in the Orszag-Tang vortex problem at $t=\pi$ obtained with (a) the HLLD scheme, (b) the MLAU scheme, (c) the MLAU scheme with the SLAU2-type pressure flux (Eq. (\ref{eq:73})). The resolution $N=200$. The reference solution with the HLLD scheme at $N=1600$ is presented in (d). The left half of the domain is shown.}
\label{fig:ot_plt}
\end{figure}

\begin{figure}
\centering
\includegraphics[clip,angle=0,scale=0.3]{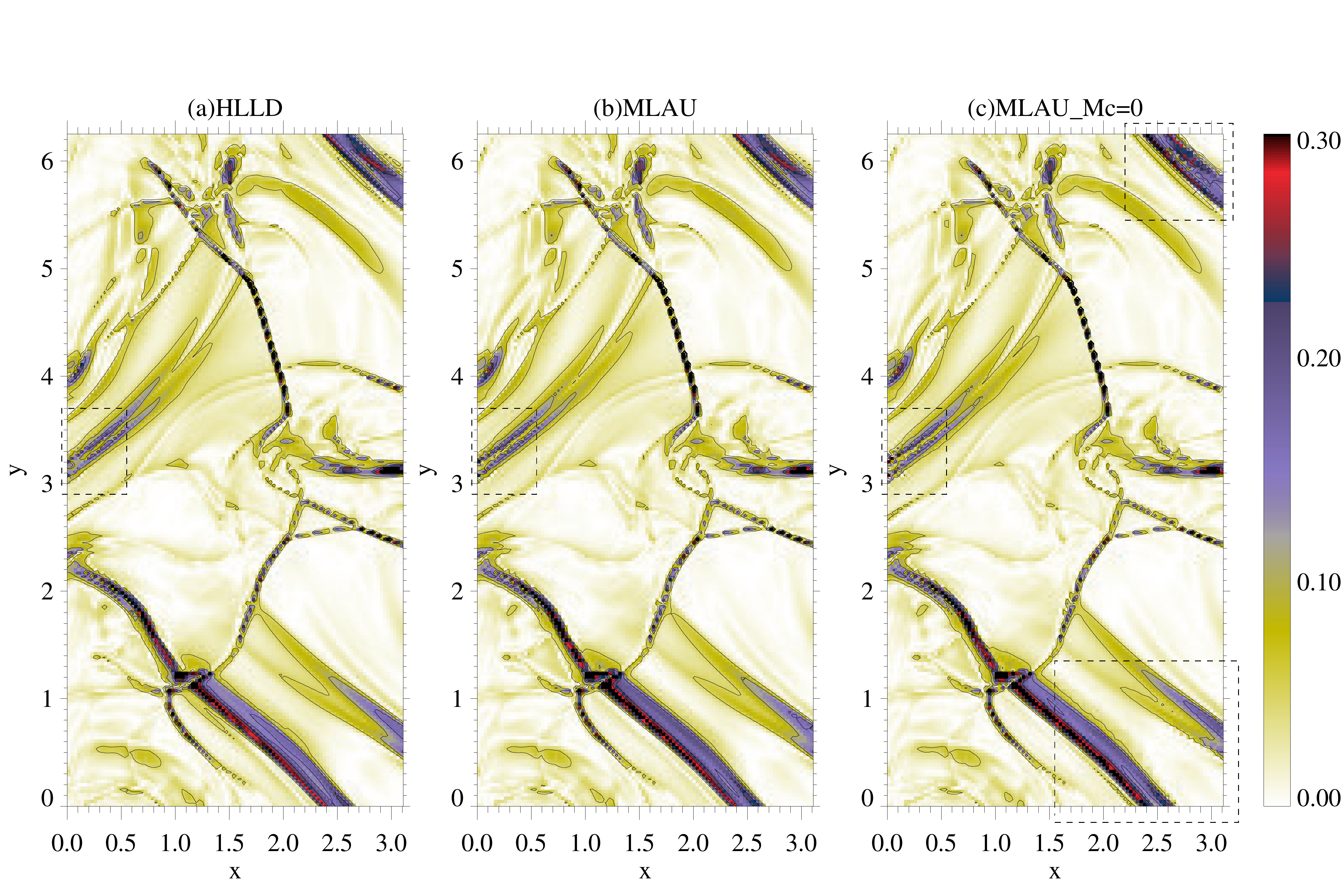}
\caption{Relative difference of the temperature profile in the Orszag-Tang vortex problem for (a) the HLLD scheme, (b) the MLAU scheme, and (c) the MLAU scheme with the SLAU2-type pressure flux (Eq. (\ref{eq:73})). Spurious oscillations are marked by dashed squares.}
\label{fig:ot_err}
\end{figure}

Figure \ref{fig:ot_plt} (a)-(b) shows the two-dimensional profile of the temperature $T=2P/\rho$ at $t=\pi$ obtained with the HLLD and MLAU schemes.
For comparison, the solution of the MLAU scheme with the SLAU2-type pressure flux (Equation (\ref{eq:73})) is shown in panel (c), and the reference solution with the HLLD scheme is presented in panel (d).
The overall structure is well reproduced in all the solutions.
Figure \ref{fig:ot_err} shows the relative difference of the temperature profile from the reference $T_{\rm ref}$, defined as $\delta T = |T-T_{\rm ref}|/T_{\rm ref}$, to examine the numerical error of solutions (a)-(c).
Noticeable numerical oscillations are marked by dashed squares.
The numerical oscillation around the current sheet at $(x,y)=(0,\pi)$ is observed in all the solutions.
Solution (c) has the largest amplitude of the error, while solution (a) exhibits the smallest.
In addition, solution (c) shows the oscillation around $(x,y)=(3.0,0.6)$ and $(x,y)=(3.0,6.0)$, at which the slow mode rarefaction wave propagates in the low plasma beta region.
This indicates that the SLAU2-type pressure flux underestimates the interface pressure for magnetized flows dominated by the slow mode, $|\vect{u}| \sim a \ll c_f$ (the third term of Equation (\ref{eq:73}) becomes negligible).
The oscillation is not seen in the solution (b), owing to the fact that the pressure flux in Equation (\ref{eq:21}) is almost uncorrected in the low beta region, $c_u \sim c_f$, and it stabilizes the slow mode wave.

\subsection{Blast wave in strongly magnetized medium}\label{sec:blast-wave-strongly}
We simulate the propagation of MHD shocks in a strongly magnetized medium, which has been tested in many literature as a simple, yet stringent problem for astrophysical phenomena \citep{1999JCoPh.149..270B,2000ApJ...530..508L,2007ApJS..170..228M,2008ApJS..178..137S,2009JCoPh.228.2480B}.
Due to the strong ambient field, the solution quality is considerably affected by the numerical errors of the magnetic field, specifically, by the error associated with the treatment of the divergence-free condition.
A family of the constrained transport method seems to provide reliable solutions \citep{2004ApJ...602.1079B,2010JCoPh.229.1970B,2012JCoPh.231.7476B,2018JCoPh.375.1365F,2019ApJS..242...14M}.
The periodic computational domain ranging from $-2<x<2$ and $-2<y<2$ is resolved by $1024\times1024$ grid points.
The initial condition $(\rho,u,v,w,B_x,B_y,B_z,P)=(1,0,0,0,B_0 \sin(\theta),B_0 \cos(\theta),0,1)$ where $B_0=10$, $\theta=30^{\circ}$, and then a high pressure cylinder is imposed at the center of the domain, $P=100$ for $\mart{x^2+y^2} < 0.125$.

\begin{figure}[htbp]
\centering
\includegraphics[clip,angle=0,scale=0.5]{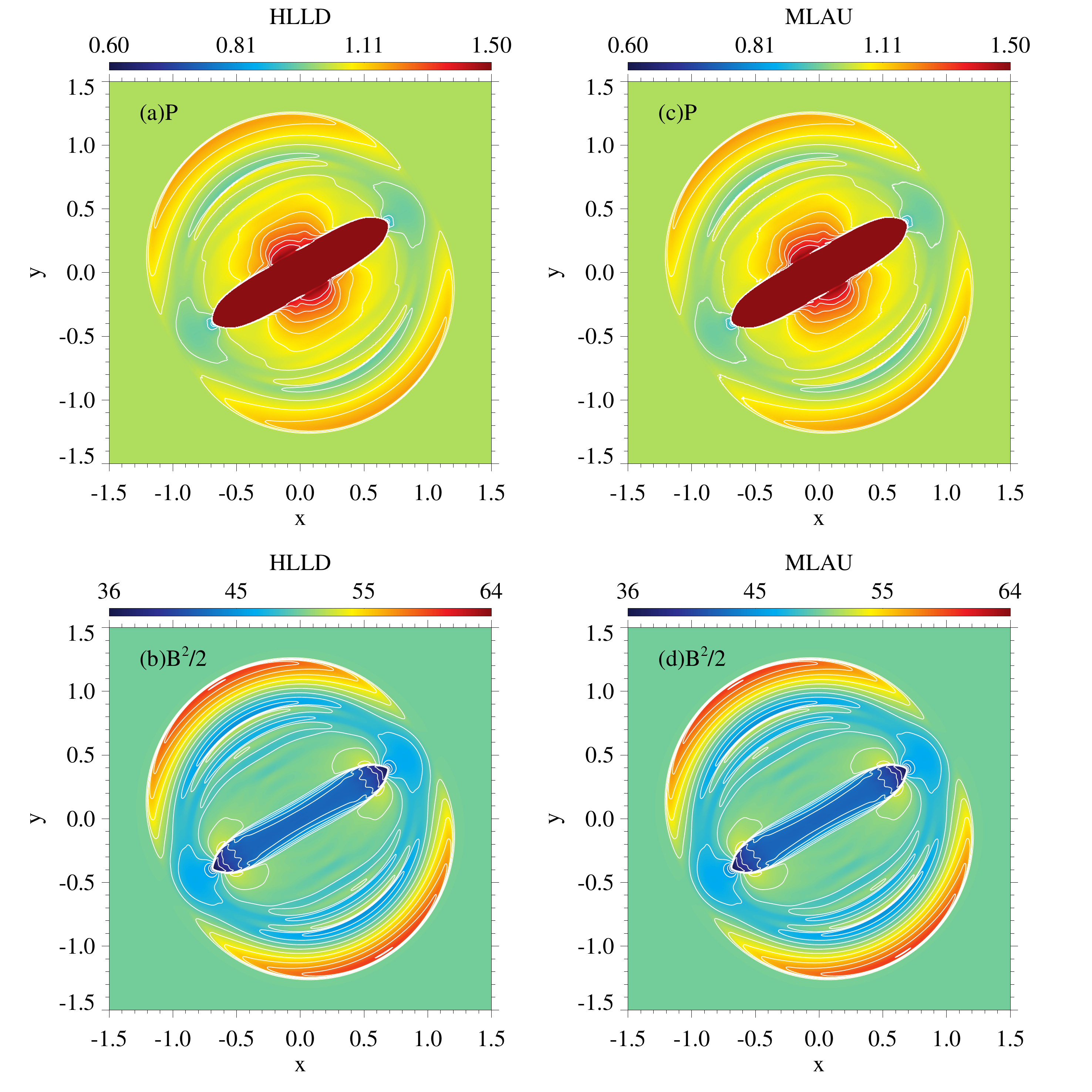}
\caption{Two-dimensional profile of the gas pressure (top) and the magnetic pressure (bottom) in the blast wave problem at $t=0.1$ obtained with the HLLD (left) and MLAU (right) schemes.}
\label{fig:blast_comp2d} 
\end{figure}

Figure \ref{fig:blast_comp2d} compares the two-dimensional profile of the gas and magnetic pressures at $t=0.1$ obtained with the HLLD (left) and MLAU (right) schemes.
Their solutions are indistinguishable from each other, and we confirm that the level of oscillation at shocks is comparable between them.
On the other hand, the Roe scheme presents negative pressure caused by the numerical error of the magnetic field, and fails to resolve this problem immediately after simulation starts.
The MLAU scheme with the SLAU2-type pressure flux (Equation (\ref{eq:73})) shows an intolerable level of oscillation around the high pressure cylinder and eventually crashes the simulation run, which stems from the deficiency of the interface pressure for strongly magnetized flows as is mentioned in Section \ref{sec:orszag-tang-vortex}.
This indicates that the numerical dissipation with a scale of {\Alfven} speed inherent in the HLLD scheme is effective for stabilizing multidimensional shocks in low beta plasma.
The MLAU scheme possesses the same level of numerical dissipation in the pressure and magnetic tension fluxes, and thus, it provides a solution of low beta plasma comparable to the HLLD scheme.

\subsection{Kelvin-Helmholtz instability}\label{sec:kelv-helmh-inst}
The Kelvin-Helmholtz instability (KHI) is the hydrodynamic instability that occurs at the velocity shear layer.
This instability can induce the mixing of different species of plasmas, for example, the entry of tailward-flowing solar wind plasma into the planetary magnetosphere through the magnetopause.
In addition, the flow can amplify the in-plane magnetic field by stretching, and it is related to the dynamo process taking place in stars, planets, and accretion disks.
We simulate the growth of the KHI to assess the capability of resolving a shear flow.
The two-dimensional computational domain ranging from $0 \leq x<L$ and $-L/2 \leq y<L/2$ is resolved by $N \times N$ cells.
The boundary condition is  periodic and symmetric in the $x$ and $y$ directions.
The initial condition has a velocity shear, $\vect{u}=((V_0/2)\tanh(y/\lambda),0,0)$, uniform density and pressure $\rho=\rho_0,P=P_0$, and a uniform magnetic field $\vect{B}=B_0(\cos\theta,0,\sin\theta)$, with $\rho_0=V_0=B_0=\lambda=1.0$ and $P_0=500$ as a fiducial run.
We use $\gamma=2.0$.
Since the Mach number of the initial flow $M_0 = 0.5V_0/a = 0.0158$, it is reasonably approximated as an incompressible flow.
To initiate the instability, we impose a small $(1\%)$ perturbation to the $y$-component of the velocity around the boundary with a wavelength equal to $L$, and the length is chosen to be the fastest growing mode under the adopted initial condition.

To verify the numerical code, we compare the linear growth rate obtained from the theory and the simulation.
The theoretical growth rate $\omega_i$ is obtained by numerically solving the linearized equations for the in-plane velocity and magnetic field, and the pressure $P_t = P+B_z^2/2$ as an eigenvalue problem as follows:
\begin{eqnarray}
\omega_{i} u   &=& -i k_x \frac{V_0}{2}\tanh\left(\frac{y}{\lambda}\right) u   - \frac{V_0}{2 \cosh^2(y/\lambda)}v   - \frac{i k_x}{\rho_0}P_t,\nonumber \\
\omega_{i} v   &=& -i k_x \frac{V_0}{2}\tanh\left(\frac{y}{\lambda}\right) v   - \frac{1}{\rho_0} \left[\pdif{P_t}{y} + B_0\cos \theta \left(\pdif{B_x}{y}-ikB_y\right)\right],\nonumber \\
\omega_{i} P_t &=& -i k_x \frac{V_0}{2}\tanh\left(\frac{y}{\lambda}\right) P_t - 2P_{t0} \left(i k_x u + \pdif{v}{y}\right),\nonumber \\
\omega_{i} B_x &=& -i k_x \frac{V_0}{2}\tanh\left(\frac{y}{\lambda}\right) B_x + \frac{V_0}{2 \cosh^2(y/\lambda)}B_y - B_0\cos\theta\pdif{v}{y},\nonumber \\
\omega_{i} B_y &=& -i k_x \frac{V_0}{2}\tanh\left(\frac{y}{\lambda}\right) B_y + ikvB_0\cos\theta,\label{eq:66}
\end{eqnarray}
where $P_{t0} = P_0+(B_0 \sin \theta)^2/2$, $k_x=2 \pi/L$, and $i$ is the imaginary unit.

\begin{figure}[htbp]
\centering
\includegraphics[clip,angle=0,scale=0.6]{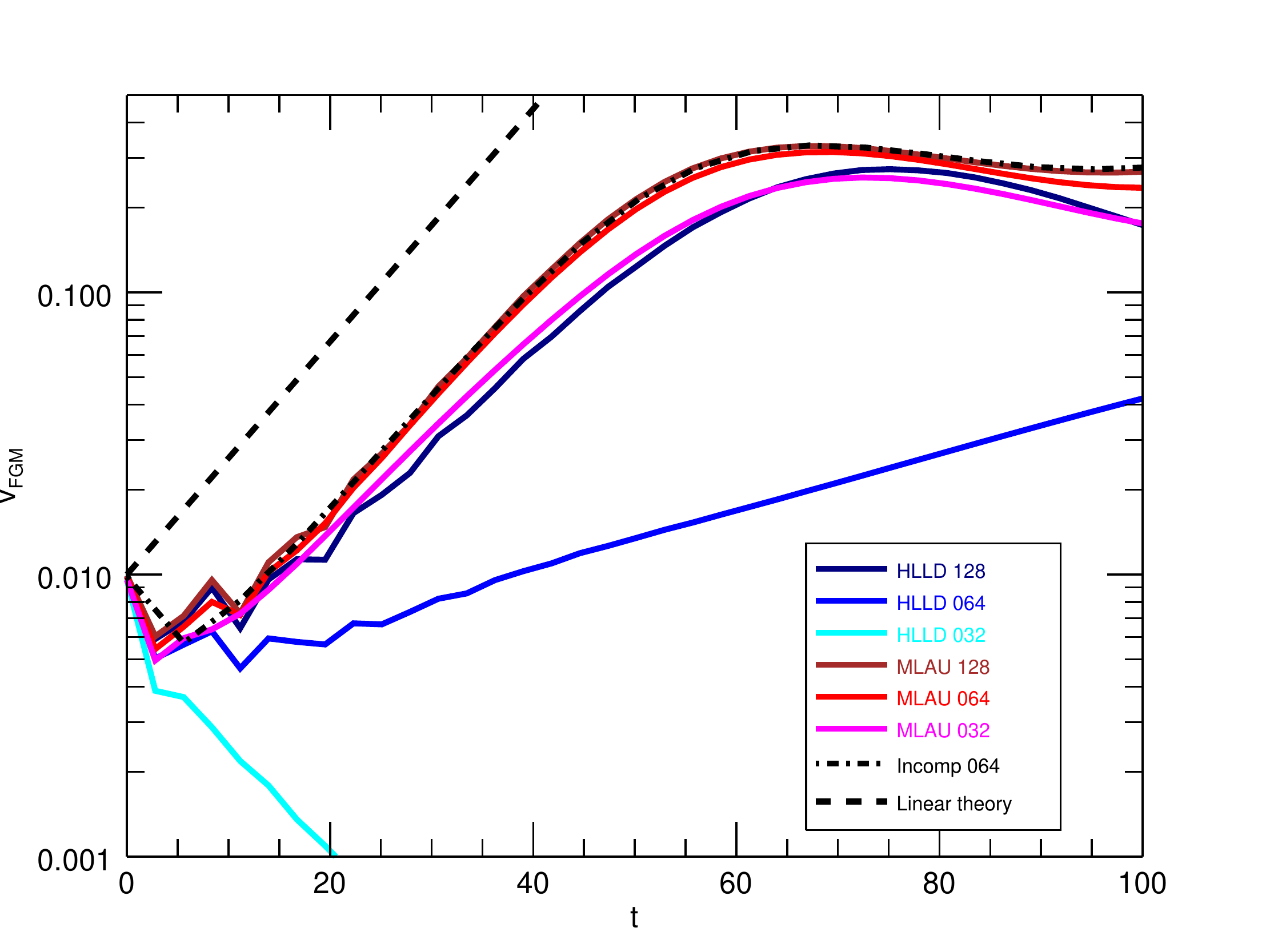}
 \caption{Time profile of the Fourier amplitude of the $y$-component of velocity in the two-dimensional Kelvin-Helmholtz instability with the out-of-plane magnetic field. The cold- and warm-colored lines are the solutions obtained with the HLLD and MLAU schemes at $N=32,64,128$. The dot-dashed line is the solution obtained from an incompressible fluid simulation with the SMAC scheme at $N=64$. The dashed line indicates the solution obtained from the linear theory.}
\label{fig:line_kh_1}
\end{figure}

The first problem is the out-of-plane magnetic field case $(\theta=90^\circ)$, $\vect{B}=(0,0,B_0)$ and $L=14\lambda$.
The situation is essentially hydrodynamic because the in-plane magnetic field is kept at zero.
Figure \ref{fig:line_kh_1} shows the growth rate obtained from the theory and the simulation.
The theoretical growth rate $\omega_{i}=0.095V_0/\lambda$ (dashed line), and the simulations are carried out at $N=32,64,128$.
Since the Mach number of the flow is very low, the familiar shock-capturing schemes suffer from excessive numerical diffusion with a scale of fast magnetosonic speed (Section \ref{sec:pressure-flux}).
As a result, the HLLD scheme cannot reproduce the theoretical growth rate at resolutions $N \leq 128$.
In contrast, the solution with the MLAU scheme converges to the theory at $N \geq 64$ and agrees with an incompressible fluid simulation result obtained with the conventional simplified marker and cell (SMAC) scheme (dot-dashed line) because of the proper scaling of the pressure flux for low Mach number flows (Equation (\ref{eq:21})).
This problem was investigated by \cite{2019ApJS..242...14M}, wherein we obtained the convergence of the HLLD scheme with a fourth-order interpolation at $N\geq 64$ (see Figure 5 therein).
Notably, the MLAU scheme is comparable to, or better than the aforementioned, despite that the scheme employs the lower-order MUSCL interpolation.

\begin{figure}[htbp]
\centering
\includegraphics[clip,angle=0,scale=0.35]{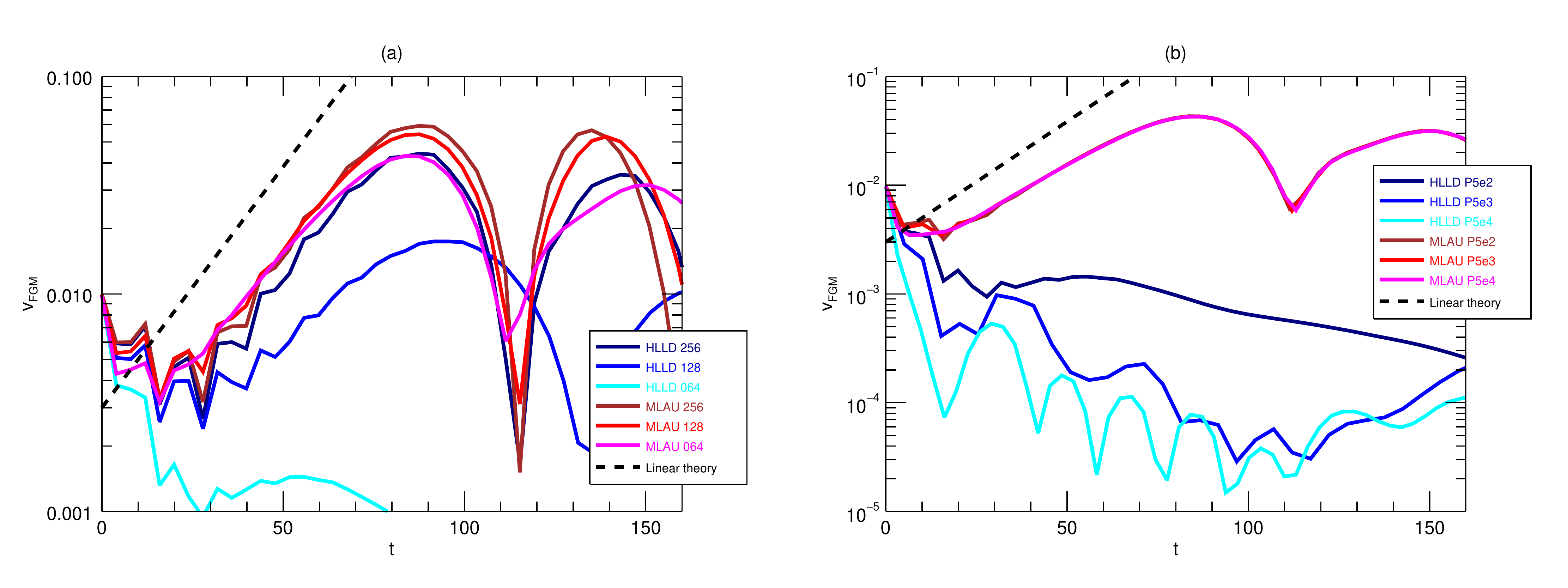}
 \caption{Same as Figure \ref{fig:line_kh_1}, but with the in-plane magnetic field. (a) Comparison among different resolutions at $N=64,128,256$. (b) Comparison among different initial pressure values $P_0=500,5000,50000$.}
\label{fig:line_mkh_uy2}
\end{figure}

The second problem entails the in-plane magnetic field $(\theta=71.565^{\circ})$, $\vect{B} = B_0(\sqrt{0.9},0,\sqrt{0.1})$ and $L=20\lambda$.
Figure \ref{fig:line_mkh_uy2}(a) shows the growth rate obtained from the theory and the simulation.
The theoretical growth rate $\omega_{i}=0.051V_0/\lambda$ and the simulations are carried out at $N=64,128,256$.
Again, the MLAU scheme converges well with the help of the pressure flux correction.
Figure \ref{fig:line_mkh_uy2}(b) compares the growth rate among different initial pressures (Mach number) $P_0=500,5000,50000 \; (M_0 = 0.0158,0.005,0.00158)$ at $N=64$.
The theoretical growth rate is identical among the three cases because the flow is almost incompressible.
While the solutions with the HLLD scheme get worse upon decreasing the Mach number, those with the MLAU scheme are independent of the Mach number.
This is indicative of the all-speed capability for MHD flows.

\begin{figure}[htbp]
\centering
\includegraphics[clip,angle=0,scale=0.4]{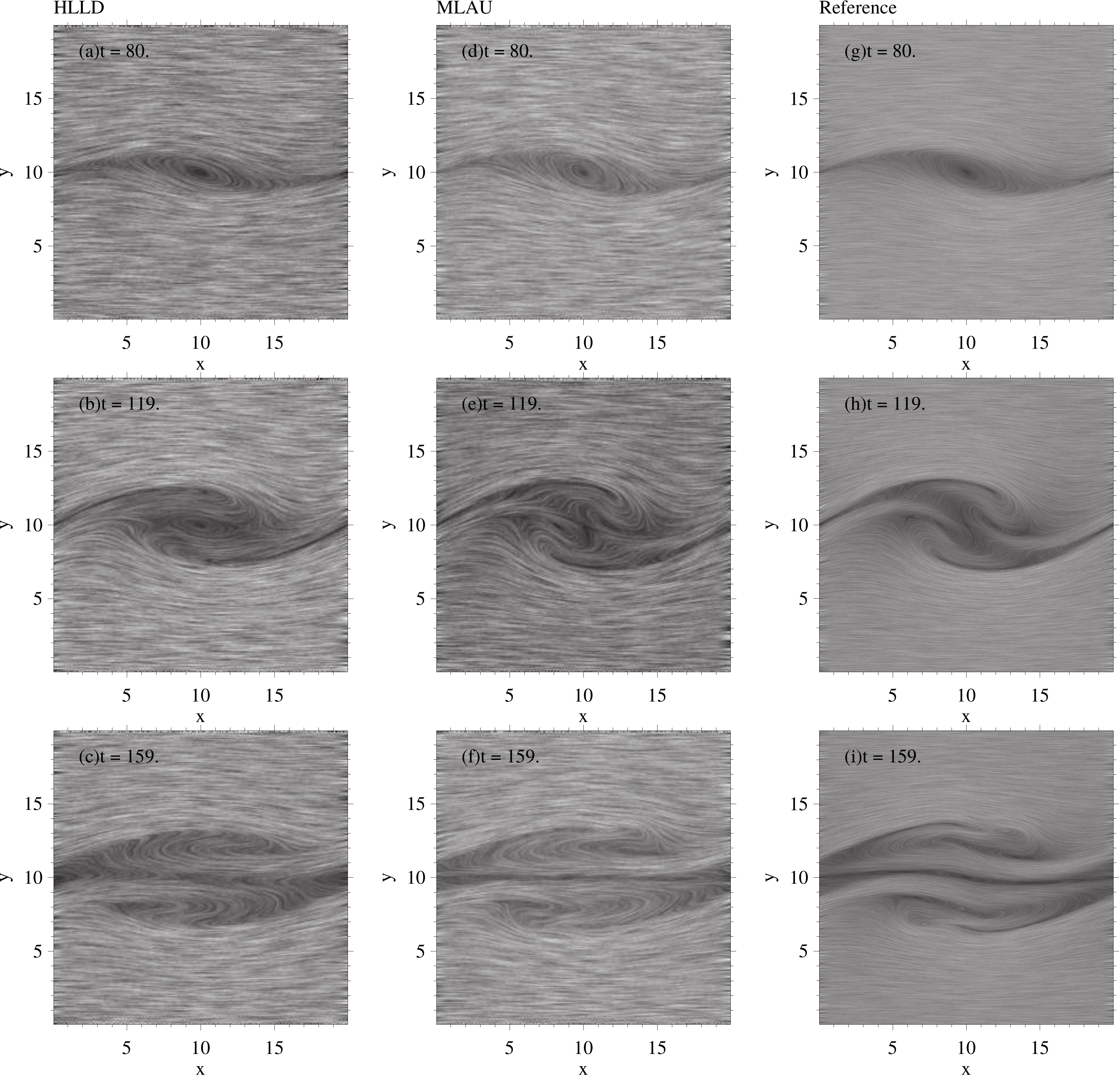}
\caption{Stream line in the Kelvin-Helmholtz instability with the in-plane magnetic field at $t=80$ (top),  $t=119$ (middle), and $t=159$ (bottom). (a)-(c) HLLD scheme at $N=256$, (d)-(f) MLAU scheme at $N=256$, and (g)-(i) HLLD scheme at $N=1024$.} 
\label{fig:plt_lic_3x3}
\end{figure}

\begin{figure}[htbp]
\centering
\includegraphics[clip,angle=0,scale=0.6]{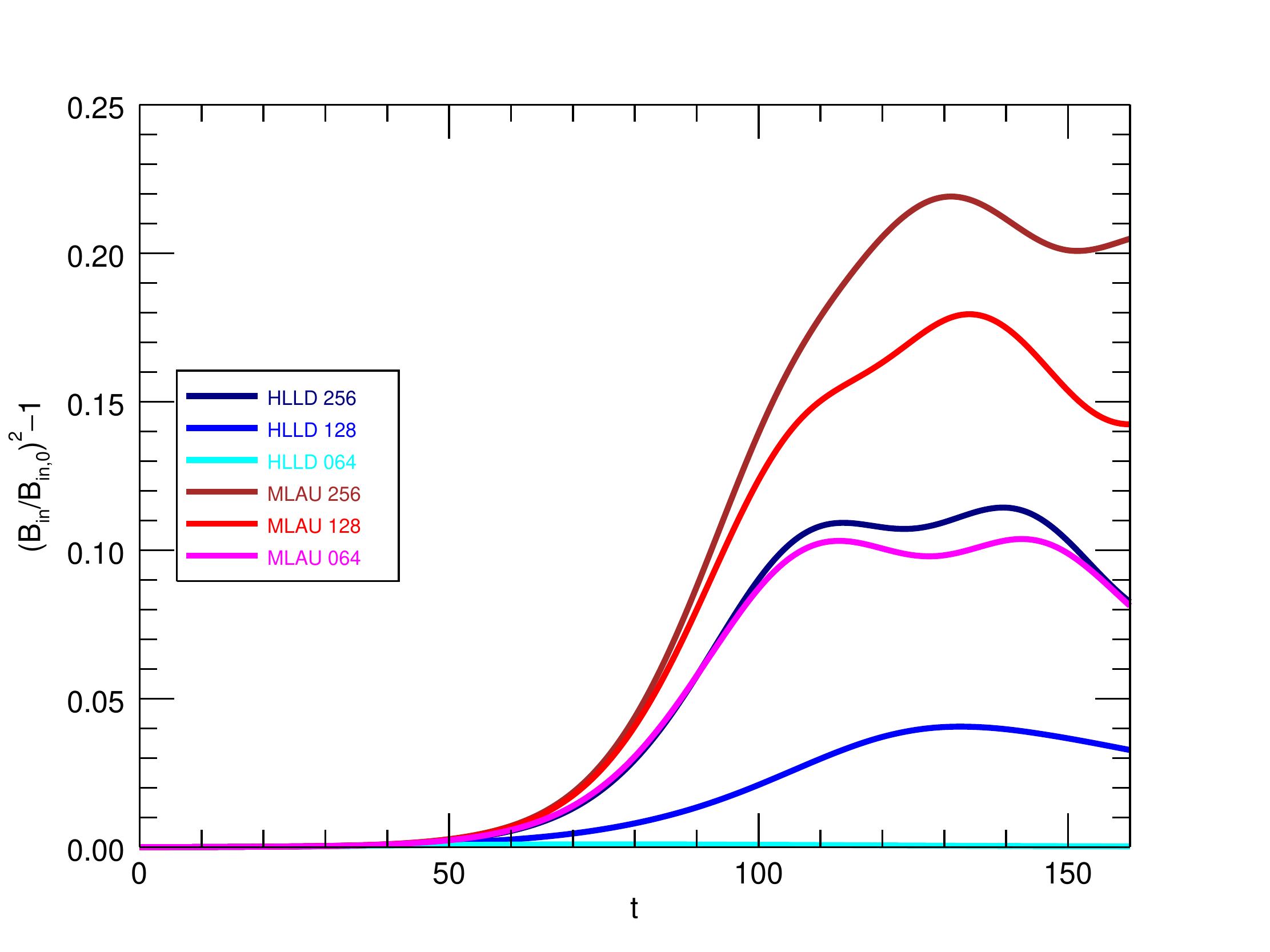}
\caption{Time profile of the increase of the in-plane magnetic field energy in the Kelvin-Helmholtz instability.} 
\label{fig:linear_mkh_bin2}
\end{figure}

Figure \ref{fig:plt_lic_3x3} shows the stream line at $t=80,119,159$ with $P_0=500$.
Panels (a)-(c) correspond to the HLLD scheme at $N=256$, (d)-(f) the MLAU scheme at $N=256$, and (g)-(i) the HLLD scheme at $N=1024$ as a reference.
At $t=119$, the flow is distorted inside the primary vortex by the magnetic tension force $(5\leq x \leq 15,7 \leq y \leq 13)$.
The MLAU scheme evidently exhibits this pattern (panel (e)), and it is in good agreement with the reference (panel (h)).
Figure \ref{fig:linear_mkh_bin2} shows the time profile of the increase of the in-plane magnetic field energy, $(B_x^2+B_y^2)/(B_x^2(0)+B_y^2(0))-1$.
The magnetic field is more amplified as the resolution is increased, as is demonstrated by \cite{2010JCoPh.229.5896M}.
The MLAU scheme achieves a higher field amplification than the HLLD scheme; in particular, it requires a quarter of the resolution to match the result with the HLLD scheme.

These problems verify that the pressure flux correction in Equation (\ref{eq:21}) greatly improves the solution accuracy, and consequently, the MLAU scheme provides a reliable solution of low Mach number MHD flows.
Note that the HLLD and MLAU schemes provide similar results for moderate Mach number flows (not shown here).

\subsection{Richtmyer-Meshkov instability}\label{sec:richtmy-meshk-inst}
When a shock collides with a corrugated contact discontinuity, the Richtmyer-Meshkov instability (RMI) is induced and the interface develops nonlinearly \citep{1960CPAM....13..297R,1972FlDy....4..101M}.
The RMI is proposed to be a mechanism of the magnetic field amplification at supernova remnants because it can amplify the magnetic field far beyond the compression by a single shock \citep{2012ApJ...758..126S,2013PhRvL.111t5001S}.
An MHD simulation of the RMI requires a sophisticated numerical scheme to robustly capture high Mach number shocks, and simultaneously, resolve the filamentary structure of the magnetic field associated with rotational flows.
Exercising caution with respect to numerical shock instability may be needed when an incident shock is well aligned with the grid spacing (e.g., planer shock propagation in Cartesian grids).
We simulate the nonlinear evolution of the RMI with the Roe, HLLD, and MLAU schemes.
The initial condition is similar to that in \cite{2019ApJS..242...14M}, i.e., $(\rho,u,v,w,B_x,B_y,B_z,P)=(1.0,0,-1.0,0,0.000034641,0,0,0.00006)$ for $y>0$ (upstream) and $(3.9988,0,-0.250075,0,0.000138523,0,0,0.75)$ for $y<0$ (downstream) that satisfies the Rankine-Hugoniot condition for perpendicular MHD shocks. 
The upstream Mach number $M=|\vect{u}|/a=100$ and the plasma beta $\beta=2P/|\vect{B}|^2=10^5$.
A corrugated contact discontinuity is imposed in the upstream region, $y_{\rm cd}=Y_{0} + \psi_0 \cos (2 \pi x/\lambda)$, where $Y_{0} = 1.0$, $\psi_0=0.1$ is a corrugation amplitude, and $\lambda=1.0$ is a wavelength.
The density increases to $\rho=10$ behind the contact discontinuity $y>y_{\rm cd}$, while \cite{2019ApJS..242...14M} investigated the case of the density decrease to $\rho=0.01$.
This setup is numerically more stringent than the previous one because the Mach number of a transmitted shock increases with decreasing sound speed.
We shift a frame moving with $v=-0.625$ that is the interface velocity after the collision with the incident shock so that the structure of the RMI stays around $y=0$ throughout the simulation run.
The domain $0 \leq x < \lambda$ and $-40 \lambda \leq y < 40 \lambda$ is resolved by $N \times 80N$ cells, where $N=128$.
The boundary condition is periodic in the $x$ direction and is fixed to be the initial state in the $y$ direction.

\begin{figure}[htbp]
\centering
\includegraphics[clip,angle=0,scale=0.2]{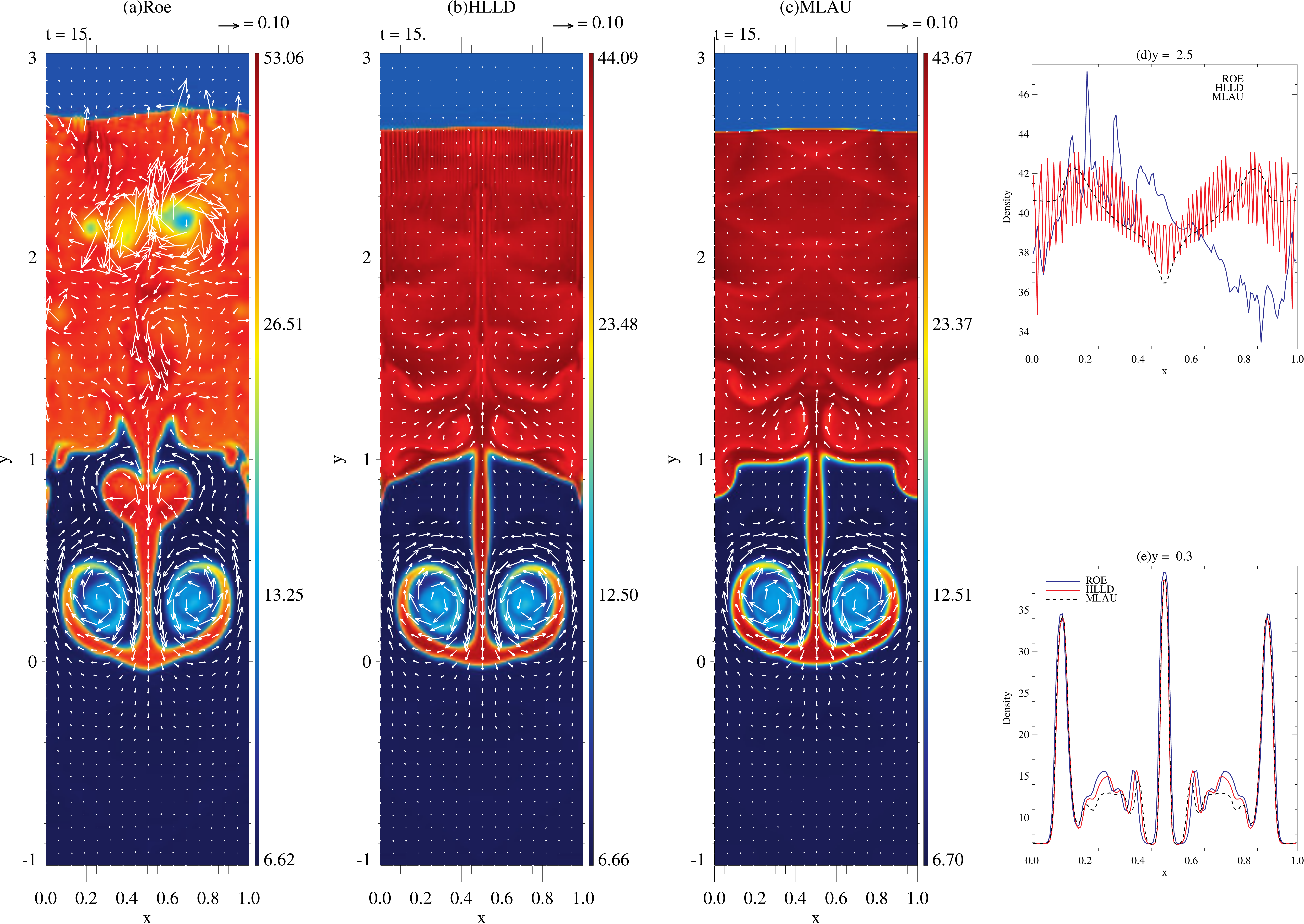}
\caption{Density profile in the Richtmyer-Meshkov instability at $t=15$. (a)-(c) Two-dimensional profile obtained with the Roe, HLLD, and MLAU schemes. Arrows represent the direction of the rotational velocity. (d)-(e) One-dimensional profiles along the shock surface at $y=2.5$ and across the spike at $y=0.3$. Blue, red, and black lines correspond to the Roe, HLLD, and MLAU schemes, respectively.}
\label{fig:rmi_plt2d_t15} 
\end{figure}

Figure \ref{fig:rmi_plt2d_t15} shows the density profile at $t=15$ obtained with the three schemes.
The surface at $y=2.6$ corresponds to the transmitted shock, and the mushroom-shaped spike in $0<y<1$ is the consequence of the nonlinear evolution of the RMI.
Apparently, the solution with the Roe scheme is severely distorted by numerical shock instability (panel (a)).
The void around $y=2.2$ was unphysically generated at the shock at $t=5$ and it flows in the $y$-direction.
The instability also damages the structure of the region of interest $(0<y<1)$, and the resulting profile deviates from the solution with the other two schemes.
Note that the simulation is carried out in the co-moving frame with the interface, which minimizes the numerical dissipation for capturing the contact discontinuity ($D^{(\rho)}$ in Equation (\ref{eq:71}) approaching zero in the $y$ direction).
The actual local Mach number $|\vect{u}|/c_f = 0.01 - 0.1$ around the spike.
This setup is a stringent condition against the numerical shock instability because the instability may be suppressed by increasing numerical dissipation for density gradient \citep{2001JCoPh.166..271P}.

\begin{figure}[htbp]
\centering
 \includegraphics[clip,angle=0,scale=0.6]{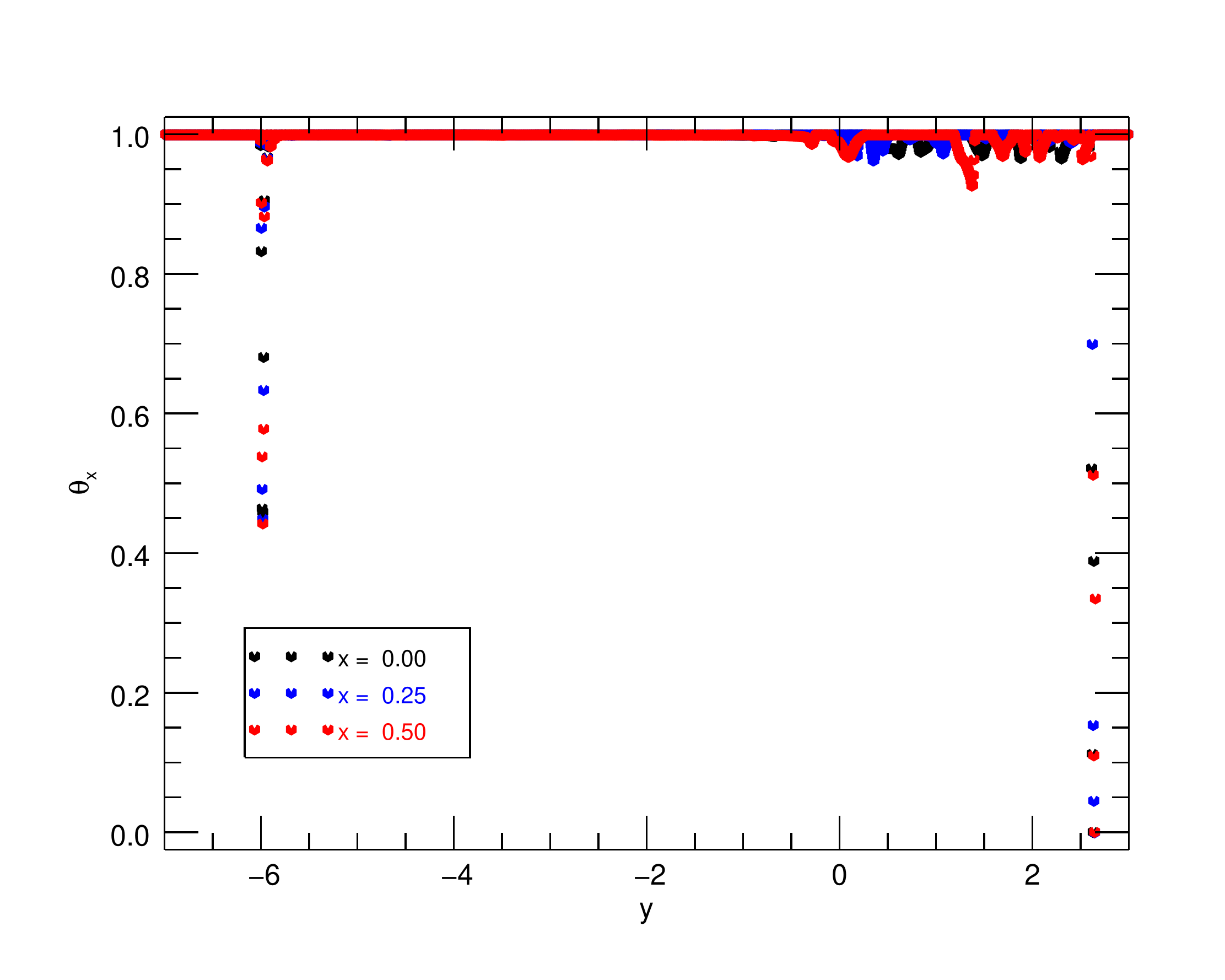}
 \caption{Shock-detecting factor $\theta$ for the $x$-component of the mass flux. The black, blue, and red symbols represent the profile at $x=0,0.25,0.5$, respectively.}
\label{fig:rmi_sd_factor} 
\end{figure}

The HLLD scheme largely remedies the catastrophic profile observed in the Roe scheme (panel (b)), but it still exhibits grid-scale oscillation at the shock front (panel (d)).
The MLAU scheme succeeds in eliminating these unphysical modes and obtains a smooth solution (panel (c)), by virtue of the shock detection in the mass flux (Equation (\ref{eq:51})).
Figure \ref{fig:rmi_sd_factor} shows the shock-detecting factor $\theta$ for the $x$-component of the mass flux.
(The factor for the $y$-component of the mass flux exceeds 0.9 over the whole domain.)
The factor $\sim 0$ at the transmitted shock at $x=2.6$, while $\sim 0.45$ at the reflected shock at $x=-6$, and $>0.9$ elsewhere; thus, it works mostly in the vicinity of shocks.
The shock detection does not degrade the accuracy of the spike, as is shown in Figure \ref{fig:rmi_plt2d_t15}(e).
When the shock detection is turned off $(\theta=1)$, the mass flux of the MLAU scheme has a finite pressure difference term in the direction parallel to the shock, and the resulting solution exhibits a grid scale oscillation similar to the HLLD scheme (not shown here).

\begin{figure}[htbp]
\centering
 \includegraphics[clip,angle=0,scale=0.6]{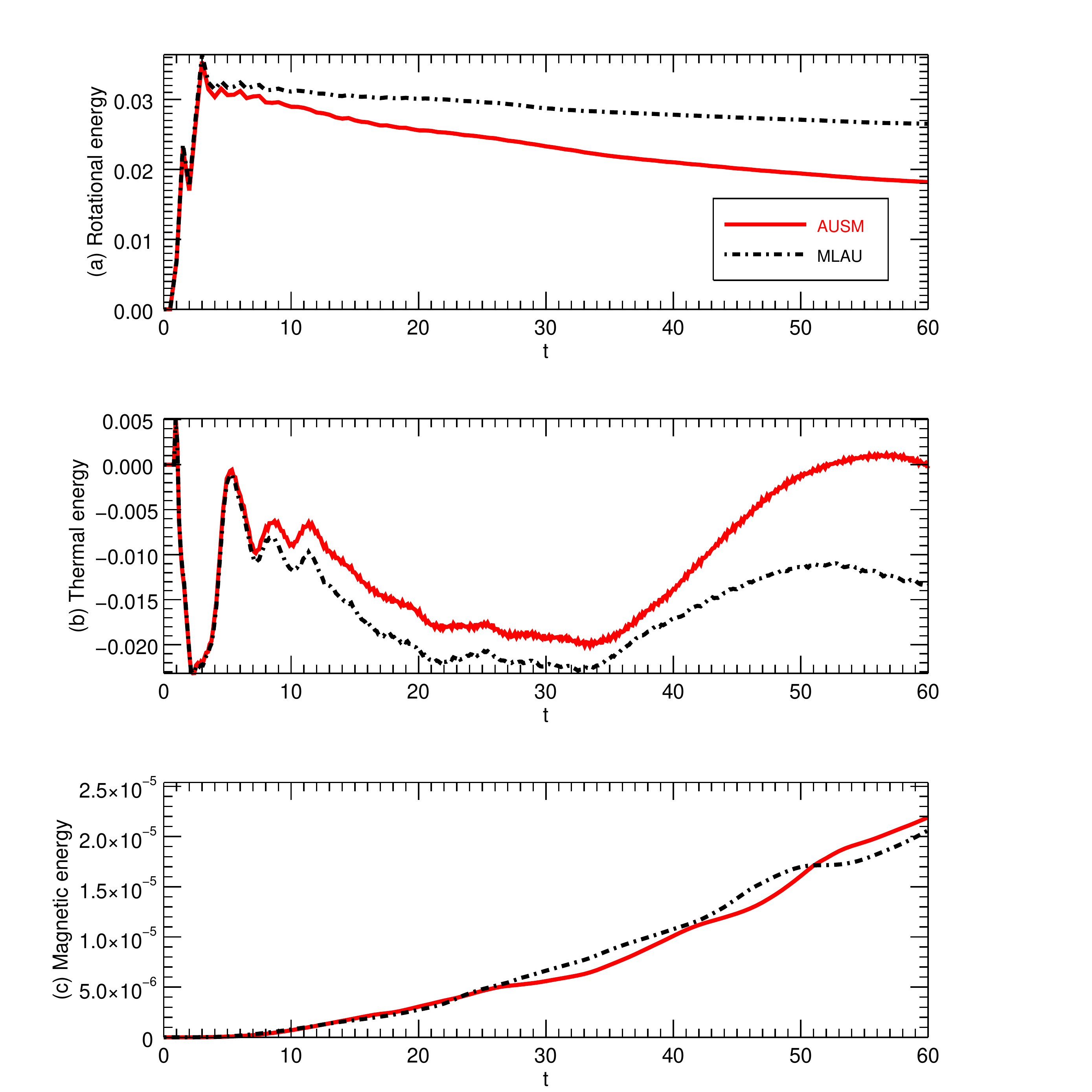}
\caption{Time profile of the increment of (a) the rotational energy $\rho |\vect{u}_R|^2/2$, (b) the thermal energy $(P-P_{\rm ref})/(\gamma-1)$, and (c) the magnetic energy $|\vect{B}|^2/2 -|\vect{B}|_{\rm ref}^2/2$ through the Richtmyer-Meshkov instability. $P_{\rm ref}$ and $\vect{B}_{\rm ref}$ are the solution in the absence of the instability. Solid red and dashed black lines are obtained with the AUSM$^{+}$-up (MLAU without the pressure flux correction) scheme and the MLAU scheme.}
\label{fig:rmi_comp3_ene}
\end{figure}

The rotational flow and the magnetic field can be amplified by the RMI, and their time evolution and saturation level are of particular interest to astrophysical application \citep{2012ApJ...758..126S,2013PhRvL.111t5001S}.
Meanwhile, their growth is sensitive to code design, such as the grid resolution, the interpolation technique, and the flux function, as well as plasma parameters \citep{2019ApJS..242...14M}.
The numerical shock instability observed in the Roe and HLLD schemes contaminates the region of interest and prevents quantitative discussion.
Consequently, we compare the solutions of the MLAU scheme with and without the pressure flux correction to check the dependence on the flux function.
The latter one, simply replacing $c_u$ in Equation (\ref{eq:21}) by $c_f$, corresponds to an MHD extension of the AUSM$^+$-up scheme.
The overall structure of the solution obtained with the AUSM$^+$-up scheme is similar to the MLAU scheme in Figure \ref{fig:rmi_plt2d_t15}(c); and thus, it is not shown here.
Figure \ref{fig:rmi_comp3_ene} shows the time profile of the increment of the spatially-averaged rotational energy $\rho |\vect{u}_R|^2/2$, thermal energy $(P-P_{\rm ref})/(\gamma-1)$, and magnetic energy $|\vect{B}|^2/2 -|\vect{B}|_{\rm ref}^2/2$ through the RMI, where $P_{\rm ref}$ and $\vect{B}_{\rm ref}$ are the solutions without corrugation.
The rotational velocity $\vect{u}_R$ satisfies $\nabla \cdot \vect{u}_R = 0$ (Equation (52) in \cite{2019ApJS..242...14M}), and is kept at zero without corrugation.
In the course of the simulation until $t=60$, the rotational energy is 1,000 times higher than the magnetic field energy; and thus, the magnetic field is continuously amplified by stretching \citep{2008PhPl...15d2102C,2012ApJ...758..126S}.
The MLAU scheme exhibits a higher level of the rotational energy and a lower level of the thermal energy than the AUSM$^+$-up scheme.
Meanwhile, the level of the magnetic energy is comparable between them.
This indicates that the rotational energy is less dissipated numerically into the thermal energy with the MLAU scheme due to the improved resolution by the pressure flux correction in the co-moving frame with the interface (third and fourth terms of Equation (\ref{eq:21}) become small).
The magnetic energy can be amplified to a level comparable to the rotational energy \citep{2019ApJS..242...14M}, although the present simulation does not reach the saturation state owing to the limited computational time and domain.
Consequently, we extrapolate the time profiles of the rotational and magnetic energies in $30<t<60$ with a linear function, and then determine the level at intersection points as a proxy for saturation.
The estimated level considerably differs according to the flux function; $7.2 \times 10^{-5}$ at $t=170$ for the AUSM$^+$-up scheme, and $2.0 \times 10^{-4}$ at $t=440$ for the MLAU scheme.

This problem undoubtedly verifies that the MLAU scheme can tackle a situation including both the low and high Mach number flows with satisfactory accuracy and stability.
The very high Mach number shock is robustly captured without the numerical instability, while the resolution of the low speed flow is improved by the pressure flux correction.

 \section{Summary}\label{sec:summary}
 We have developed a novel numerical scheme for ideal magnetohydrodynamic (MHD) simulations based on the AUSM scheme \citep{1993JCoPh.107...23L,1996JCoPh.129..364L}, which is a simple, yet accurate and robust scheme employed in computational aerodynamics.
The present scheme splits the inviscid flux in MHD equations into the advection, pressure, and magnetic tension parts, and then individually evaluates the mass, pressure, and magnetic tension fluxes.
The mass flux follows the AUSM$^+$-up scheme \citep{2006JCoPh.214..137L} upon including shock detection to control the pressure difference term to avoid numerical shock instability in multidimension.
The magnetic tension flux is derived to be consistent with the HLLD approximate Riemann solver \citep{2005JCoPh.208..315M}.
As opposed to preceding AUSM-family schemes for MHD simulations, the present scheme can correctly capture MHD tangential and rotational discontinuities, as well as the contact discontinuity.
The pressure flux is a weighted average of the left and right states similar to the shock-capturing schemes, and is then corrected to improve the resolution of low Mach number flows in high beta plasma.
In low beta plasma, on the other hand, the pressure flux is comparable to that of the HLLD scheme, and it provides numerical dissipation sufficient to stabilize magnetized flows.

We measure the performance of the present scheme through various benchmark tests, including shocks, waves, and low and high Mach number flows.
In standard benchmark tests (shock problems in Sections \ref{sec:one-dimens-shock} and \ref{sec:blast-wave-strongly}, and Orszag-Tang vortex in Section \ref{sec:orszag-tang-vortex}), the solution with the present scheme is comparable to the HLLD scheme, and it is more robust than the Roe scheme.
We verify the advantages of the present scheme through stringent problems.
In the problem of the low Mach number MHD Kelvin-Helmholtz instability (Section \ref{sec:kelv-helmh-inst}), the solution with the present scheme is independent of the Mach number and agrees with the linear theory and the incompressible solution (``all-speed'' capability), while the familiar shock-capturing scheme gets worse upon decreasing the Mach number.
This advantage stems from the pressure flux with a scale of advection speed rather than fast magnetosonic speed (Equation (\ref{eq:21})).
We expect that the present scheme maintains all-speed capability as long as the flow is approximately faster than the {\Alfven} speed because the pressure flux properly scales down to $|\vect{u}|=c_a$ and the magnetic tension flux has a numerical diffusion term with a scale of {\Alfven} speed inherited from the baseline HLLD scheme that will suppress sub-{\Alfven}ic flows.
Therefore, we consider that the present scheme achieves ``{\it quasi}-all-speed'' capability for super-{\Alfven}ic flows.

The problem of the Richtmyer-Meshkov instability (RMI) involves both high and low Mach number flows (Section \ref{sec:richtmy-meshk-inst}).
The present scheme can capture a very high Mach number shock without numerical instability, such as the odd-even decoupling and the Carbuncle phenomena found in familiar shock-capturing schemes, by virtue of the shock detection in the mass flux.
Given that the shock detection solely drops the pressure difference term in the direction parallel to the shock that is a possible cause of the numerical instability \citep{2000JCoPh.160..623L}, it does not degrade the accuracy of the contact discontinuity, which is of particular interest for the RMI.
As a consequence of the RMI, rotational flows grow and the magnetic field is amplified there.
The evolution of the rotational energy in the nonlinear stage depends on the resolution of the flux function.
The present scheme with the pressure flux correction exhibits higher levels of rotational energy than the scheme without the correction due to the reduced numerical dissipation.
It may provide higher saturation levels of the magnetic energy.

Since the present work employs an explicit time integration method, the time step of the simulation is restricted by the CFL condition of the propagation speed of the fast mode.
Thus, it requires a huge number of time steps to solve the dynamics of convective motion in nearly incompressible flows.
Owing to the CFL restriction, the explicit method is also poor at solving problems using local fine grid spacing, e.g., the boundary layer problem.
An implicit time integration method is frequently employed in computational aerodynamics to deal with these issues, but the implicit method is not necessarily suited for modern massively-parallel supercomputers.
\cite{2012A&A...539A..30H} employed the reduced speed of sound technique to relax the CFL restriction and reduce the computational cost of low Mach number flow simulations with an explicit scheme.
Recently, \cite{2019A&A...622A.157I} modified the technique to tackle low Mach number flows with a large density variation.
In the new technique, the basic equations are written in a semiconservative form, and they reduce to a conservative form when the sound speed is unchanged.
Therefore, it is applicable to both low and high Mach number flows.
Combining this technique with the present scheme might enhance both the accuracy of the solution and the computational efficiency of simulations of wide-ranging Mach number flows.

Through the benchmark tests, we verify that the present scheme satisfies the following three properties, namely, (i) robust capturing of one- and multi-dimensional shocks, (ii) accurate solution of low Mach number flows, i.e., quasi-all-speed capability for super-{\Alfven}ic flows, and (iii) correct capturing of MHD discontinuities.
Hence, the present scheme must be a promising tool for MHD simulations of space and astrophysical plasmas that include both low and high Mach number flows, shocks, turbulence, and magnetic field inhomogeneities, for example, the solar system from the Sun's interior and atmosphere, interplanetary space, planetary magnetosphere, termination shock, and beyond.

\begin{acknowledgements}
We thank the anonymous referee for carefully reviewing our manuscript and giving insightful comments.
We thank T. Hanawa, Y. Matsumoto, H. Hotta, T. Mamashita, and R. Matsumoto for their fruitful discussion.
We also thank H. Iijima for reading our manuscript and giving comments.
K. Kitamura was partially supported by JSPS KAKENHI grand Number JP19K04834.
We would like to thank Editage (www.editage.com) for English language editing.
\end{acknowledgements}

\appendix
\section{Hyperbolic System of Preconditioned Euler Equations}\label{sec:prec-syst-euler}
The one-dimensional Euler equations are rewritten for symmetrizing variables $d\vect{Q}=[dS,du,dP/\rho c]^T, dS = dP-a^2d\rho$ as:
\begin{eqnarray}
 \pdif{\vect{Q}}{t} + \left(\vect{L} \vect{A} \vect{R}\right) \pdif{\vect{Q}}{x}= 0,\label{eq:58}
\end{eqnarray}
where,
\begin{eqnarray}
\vect{A} = 
\begin{bmatrix}
 u& \rho& 0\\
0 & u & \frac{1}{\rho}\\
0 & \rho a^2 & u
\end{bmatrix}
, \vect{L} =
\begin{bmatrix}
 -a^2 & 0 & 1\\
0 & 1 & 0 \\
0 & 0 & \frac{1}{\rho a}
\end{bmatrix}
, \vect{R} = 
\begin{bmatrix}
 -\frac{1}{a^2} & 0 & \frac{\rho}{a}\\
0 & 1 & 0 \\
0 & 0 & \rho a
\end{bmatrix}
,\label{eq:60}
\end{eqnarray}
and $(\vect{L} \vect{A} \vect{R})$  is symmetric.
Applying a preconditioning matrix $\vect{\Gamma}^{-1} = {\rm diag}(1,1,\epsilon)$ proposed by \cite{1995AIAAJ..33.2050W}, the preconditioned Jacobian expressed in primitive variables $(\rho,u,P)$ is obtained as:
\begin{eqnarray}
 \tilde{\vect{A}} = \vect{R} \left(\vect{\Gamma}^{-1} \vect{L} \vect{A} \vect{R} \right) \vect{L} = 
\begin{bmatrix}
u & \epsilon \rho& -(1-\epsilon) \frac{u}{a^2}\\
0 & u & \frac{1}{\rho} \\
0 & \epsilon \rho a^2 & \epsilon u
\end{bmatrix}
,\label{eq:61}
\end{eqnarray}
and its eigenvalues and left- and right-eigenvectors are:
\begin{eqnarray}
&& \lambda_1 = u, \; \vect{l}_1 = (1,0,-1/a^2), \; \vect{r}_1 = (1,0,0)^T,\label{eq:62}\\
&& \lambda_2 = u - a_+, \; \vect{l}_2 = \left(0,-\frac{\rho a_+ a_-}{(a_+ + a_-)a^2},\frac{a_+}{(a_+ + a_-)a^2}\right), \; \vect{r}_2 = \left(1,-\frac{a^2}{\rho a_+},a^2\right)^T,\label{eq:63}\\
&& \lambda_3 = u + a_-, \; \vect{l}_3 = \left(0, \frac{\rho a_+ a_-}{(a_+ + a_-)a^2},\frac{a_-}{(a_+ + a_-)a^2}\right), \; \vect{r}_3 = \left(1, \frac{a^2}{\rho a_-},a^2\right)^T,\label{eq:64}
\end{eqnarray}
where,
\begin{eqnarray}
 a_{\pm} = \frac{\left[ \sqrt{(1-\epsilon)^2 u^2 + 4 \epsilon a^2} \pm (1-\epsilon)u\right]}{2} = \left\{
\begin{array}{l}
 a, \;\;\; {\rm if} \; \epsilon=1, \\
(\sqrt{5}|u| \pm u)/2, \;\;\; {\rm if} \; \epsilon = u^2/a^2 \ll 1,
\end{array}
\right.\label{eq:65}
\end{eqnarray}
modifies the wave speed to be a scale of advection speed if $\epsilon \sim M^2$.
From the characteristic relations $\vect{l} \cdot d\vect{Q} = 0$,
\begin{eqnarray}
\left\{
\begin{array}{c}
 dP - \rho a_- du = 0, \;\;\; {\rm along} \; \lambda_2,\\
 dP + \rho a_+ du = 0, \;\;\; {\rm along} \; \lambda_3.
\end{array}
\right.\label{eq:69}
\end{eqnarray}
Discretizing them along characteristics ($\lambda_2<0,\lambda_3>0$), we approximate the interface pressure as:
\begin{eqnarray}
P_{1/2} &=& \frac{1}{2} \left[P_L + P_R - \rho_{1/2}a\Delta u \right] \;\;\; {\rm if} \; \epsilon = 1 \label{eq:74} \\ 
&=& \frac{1}{2} \left[P_L + P_R - \frac{\sqrt{5}}{2}\rho_{1/2}|u_{1/2}|\Delta u \right] \;\;\; {\rm if} \; \epsilon = u^2/a^2 \ll 1.\label{eq:68}
\end{eqnarray}
The velocity difference term in Equation (\ref{eq:68}) has the same scaling as the pressure flux corrected in Section \ref{sec:pressure-flux}, while the term originally scales as the speed of sound in Equation (\ref{eq:74}).


\begin{thebibliography}{}
\expandafter\ifx\csname natexlab\endcsname\relax\def\natexlab#1{#1}\fi
\providecommand{\url}[1]{\href{#1}{#1}}
\providecommand{\dodoi}[1]{doi:~\href{http://doi.org/#1}{\nolinkurl{#1}}}
\providecommand{\doeprint}[1]{\href{http://ascl.net/#1}{\nolinkurl{http://ascl.net/#1}}}
\providecommand{\doarXiv}[1]{\href{https://arxiv.org/abs/#1}{\nolinkurl{https://arxiv.org/abs/#1}}}

\bibitem[{{Balsara}(1998)}]{1998ApJS..116..119B}
{Balsara}, D.~S. 1998, \apjs, 116, 119, \dodoi{10.1086/313092}

\bibitem[{{Balsara}(2010)}]{2010JCoPh.229.1970B}
---. 2010, Journal of Computational Physics, 229, 1970,
  \dodoi{10.1016/j.jcp.2009.11.018}

\bibitem[{{Balsara}(2012)}]{2012JCoPh.231.7476B}
---. 2012, Journal of Computational Physics, 231, 7476,
  \dodoi{10.1016/j.jcp.2011.12.025}

\bibitem[{{Balsara} \& {Kim}(2004)}]{2004ApJ...602.1079B}
{Balsara}, D.~S., \& {Kim}, J. 2004, \apj, 602, 1079, \dodoi{10.1086/381051}

\bibitem[{{Balsara} {et~al.}(2009){Balsara}, {Rumpf}, {Dumbser}, \&
  {Munz}}]{2009JCoPh.228.2480B}
{Balsara}, D.~S., {Rumpf}, T., {Dumbser}, M., \& {Munz}, C.-D. 2009, Journal of
  Computational Physics, 228, 2480, \dodoi{10.1016/j.jcp.2008.12.003}

\bibitem[{{Balsara} \& {Spicer}(1999)}]{1999JCoPh.149..270B}
{Balsara}, D.~S., \& {Spicer}, D.~S. 1999, Journal of Computational Physics,
  149, 270, \dodoi{10.1006/jcph.1998.6153}

\bibitem[{{Brio} \& {Wu}(1988)}]{1988JCoPh..75..400B}
{Brio}, M., \& {Wu}, C.~C. 1988, Journal of Computational Physics, 75, 400,
  \dodoi{10.1016/0021-9991(88)90120-9}

\bibitem[{{Cao} {et~al.}(2008){Cao}, {Wu}, {Ren}, \&
  {Li}}]{2008PhPl...15d2102C}
{Cao}, J., {Wu}, Z., {Ren}, H., \& {Li}, D. 2008, Physics of Plasmas, 15,
  042102, \dodoi{10.1063/1.2842367}

\bibitem[{{Dai} \& {Woodward}(1994)}]{1994JCoPh.111..354D}
{Dai}, W., \& {Woodward}, P.~R. 1994, Journal of Computational Physics, 111,
  354, \dodoi{10.1006/jcph.1994.1069}

\bibitem[{{Felker} \& {Stone}(2018)}]{2018JCoPh.375.1365F}
{Felker}, K.~G., \& {Stone}, J.~M. 2018, Journal of Computational Physics, 375,
  1365, \dodoi{10.1016/j.jcp.2018.08.025}

\bibitem[{{Fromang} {et~al.}(2006){Fromang}, {Hennebelle}, \&
  {Teyssier}}]{2006A&A...457..371F}
{Fromang}, S., {Hennebelle}, P., \& {Teyssier}, R. 2006, \aap, 457, 371,
  \dodoi{10.1051/0004-6361:20065371}

\bibitem[{{Godunov}(1959)}]{1959GODUNOV}
{Godunov}, S.~K. 1959, Matematicheskii Sbornik, 47, 271

\bibitem[{{Gottlieb} \& {Shu}(1998)}]{1998MaCom..67...73G}
{Gottlieb}, S., \& {Shu}, C.~W. 1998, Mathematics of Computation, 67, 73

\bibitem[{{Han} {et~al.}(2009){Han}, {Lee}, \& {Kim}}]{2009AIAAJ..47..970H}
{Han}, S.-H., {Lee}, J.-I., \& {Kim}, K.~H. 2009, AIAA Journal, 47, 970,
  \dodoi{10.2514/1.39375}

\bibitem[{{Hanawa} {et~al.}(2008){Hanawa}, {Mikami}, \&
  {Matsumoto}}]{2008JCoPh.227.7952H}
{Hanawa}, T., {Mikami}, H., \& {Matsumoto}, T. 2008, Journal of Computational
  Physics, 227, 7952, \dodoi{10.1016/j.jcp.2008.05.006}

\bibitem[{{Harten} {et~al.}(1983){Harten}, {Lax}, \& {van
  Leer}}]{1983SIAMrev.25..35M}
{Harten}, A., {Lax}, P., \& {van Leer}, B. 1983, SIAM review, 35, 35

\bibitem[{{Hotta} {et~al.}(2012){Hotta}, {Rempel}, {Yokoyama}, {Iida}, \&
  {Fan}}]{2012A&A...539A..30H}
{Hotta}, H., {Rempel}, M., {Yokoyama}, T., {Iida}, Y., \& {Fan}, Y. 2012, \aap,
  539, A30, \dodoi{10.1051/0004-6361/201118268}

\bibitem[{{Iijima} {et~al.}(2019){Iijima}, {Hotta}, \&
  {Imada}}]{2019A&A...622A.157I}
{Iijima}, H., {Hotta}, H., \& {Imada}, S. 2019, \aap, 622, A157,
  \dodoi{10.1051/0004-6361/201834031}

\bibitem[{{Kim} {et~al.}(2003){Kim}, {Kim}, {Rho}, \& {Kyu
  Hong}}]{2003JCoPh.185..342K}
{Kim}, S.-s., {Kim}, C., {Rho}, O.-H., \& {Kyu Hong}, S. 2003, Journal of
  Computational Physics, 185, 342, \dodoi{10.1016/S0021-9991(02)00037-2}

\bibitem[{{Kitamura} \& {Balsara}(2019)}]{2019ShWav..29..611K}
{Kitamura}, K., \& {Balsara}, D.~S. 2019, Shock Waves, 29, 611,
  \dodoi{10.1007/s00193-018-0842-0}

\bibitem[{{Kitamura} \& {Shima}(2013)}]{2013JCoPh.245...62K}
{Kitamura}, K., \& {Shima}, E. 2013, Journal of Computational Physics, 245, 62,
  \dodoi{10.1016/j.jcp.2013.02.046}

\bibitem[{{Kritsuk} {et~al.}(2011){Kritsuk}, {Nordlund}, {Collins}, {Padoan},
  {Norman}, {Abel}, {Banerjee}, {Federrath}, {Flock}, {Lee}, {Li},
  {M{\"u}ller}, {Teyssier}, {Ustyugov}, {Vogel}, \& {Xu}}]{2011ApJ...737...13K}
{Kritsuk}, A.~G., {Nordlund}, {\AA}., {Collins}, D., {et~al.} 2011, \apj, 737,
  13, \dodoi{10.1088/0004-637X/737/1/13}

\bibitem[{{Lee} \& {Deane}(2009)}]{2009JCoPh.228..952L}
{Lee}, D., \& {Deane}, A.~E. 2009, Journal of Computational Physics, 228, 952,
  \dodoi{10.1016/j.jcp.2008.08.026}

\bibitem[{{Lesur} \& {Longaretti}(2005)}]{2005A&A...444...25L}
{Lesur}, G., \& {Longaretti}, P.~Y. 2005, \aap, 444, 25,
  \dodoi{10.1051/0004-6361:20053683}

\bibitem[{{Li}(2005)}]{2005JCoPh.203..344L}
{Li}, S. 2005, Journal of Computational Physics, 203, 344,
  \dodoi{10.1016/j.jcp.2004.08.020}

\bibitem[{{Liou}(1996)}]{1996JCoPh.129..364L}
{Liou}, M.-S. 1996, Journal of Computational Physics, 129, 364,
  \dodoi{10.1006/jcph.1996.0256}

\bibitem[{{Liou}(2000)}]{2000JCoPh.160..623L}
---. 2000, Journal of Computational Physics, 160, 623,
  \dodoi{10.1006/jcph.2000.6478}

\bibitem[{{Liou}(2006)}]{2006JCoPh.214..137L}
---. 2006, Journal of Computational Physics, 214, 137,
  \dodoi{10.1016/j.jcp.2005.09.020}

\bibitem[{{Liou} \& {Steffen}(1993)}]{1993JCoPh.107...23L}
{Liou}, M.-S., \& {Steffen}, C.~J. 1993, Journal of Computational Physics, 107,
  23, \dodoi{10.1006/jcph.1993.1122}

\bibitem[{{Londrillo} \& {Del Zanna}(2000)}]{2000ApJ...530..508L}
{Londrillo}, P., \& {Del Zanna}, L. 2000, \apj, 530, 508,
  \dodoi{10.1086/308344}

\bibitem[{{Maron} \& {Goldreich}(2001)}]{2001ApJ...554.1175M}
{Maron}, J., \& {Goldreich}, P. 2001, \apj, 554, 1175, \dodoi{10.1086/321413}

\bibitem[{{Matsumoto} {et~al.}(2019){Matsumoto}, {Asahina}, {Kudoh},
  {Kawashima}, {Matsumoto}, {Takahashi}, {Minoshima}, {Zenitani}, {Miyoshi}, \&
  {Matsumoto}}]{2019PASJ...71...83M}
{Matsumoto}, Y., {Asahina}, Y., {Kudoh}, Y., {et~al.} 2019, \pasj, 71, 83,
  \dodoi{10.1093/pasj/psz064}

\bibitem[{{Meshkov}(1972)}]{1972FlDy....4..101M}
{Meshkov}, E.~E. 1972, Fluid Dynamics, 4, 101, \dodoi{10.1007/BF01015969}

\bibitem[{{Mignone} {et~al.}(2007){Mignone}, {Bodo}, {Massaglia}, {Matsakos},
  {Tesileanu}, {Zanni}, \& {Ferrari}}]{2007ApJS..170..228M}
{Mignone}, A., {Bodo}, G., {Massaglia}, S., {et~al.} 2007, \apjs, 170, 228,
  \dodoi{10.1086/513316}

\bibitem[{{Mignone} {et~al.}(2010){Mignone}, {Tzeferacos}, \&
  {Bodo}}]{2010JCoPh.229.5896M}
{Mignone}, A., {Tzeferacos}, P., \& {Bodo}, G. 2010, Journal of Computational
  Physics, 229, 5896, \dodoi{10.1016/j.jcp.2010.04.013}

\bibitem[{{Minoshima} {et~al.}(2019){Minoshima}, {Miyoshi}, \&
  {Matsumoto}}]{2019ApJS..242...14M}
{Minoshima}, T., {Miyoshi}, T., \& {Matsumoto}, Y. 2019, \apjs, 242, 14,
  \dodoi{10.3847/1538-4365/ab1a36}

\bibitem[{{Miyoshi} \& {Kusano}(2005)}]{2005JCoPh.208..315M}
{Miyoshi}, T., \& {Kusano}, K. 2005, Journal of Computational Physics, 208,
  315, \dodoi{10.1016/j.jcp.2005.02.017}

\bibitem[{{Nishikawa} \& {Kitamura}(2008)}]{2008JCoPh.227.2560N}
{Nishikawa}, H., \& {Kitamura}, K. 2008, Journal of Computational Physics, 227,
  2560, \dodoi{10.1016/j.jcp.2007.11.003}

\bibitem[{{Orszag} \& {Tang}(1979)}]{1979JFM....90..129O}
{Orszag}, S.~A., \& {Tang}, C.-M. 1979, Journal of Fluid Mechanics, 90, 129,
  \dodoi{10.1017/S002211207900210X}

\bibitem[{{Pandolfi} \& {D'Ambrosio}(2001)}]{2001JCoPh.166..271P}
{Pandolfi}, M., \& {D'Ambrosio}, D. 2001, Journal of Computational Physics,
  166, 271, \dodoi{10.1006/jcph.2000.6652}

\bibitem[{{Qiu} \& {Shu}(2002)}]{2002JCoPh.183..187Q}
{Qiu}, J., \& {Shu}, C.-W. 2002, Journal of Computational Physics, 183, 187,
  \dodoi{10.1006/jcph.2002.7191}

\bibitem[{{Quirk}(1994)}]{1994IJNMF..18..555Q}
{Quirk}, J.~J. 1994, International Journal for Numerical Methods in Fluids, 18,
  555, \dodoi{10.1002/fld.1650180603}

\bibitem[{{Richtmyer}(1960)}]{1960CPAM....13..297R}
{Richtmyer}, R.~D. 1960, Communications on Pure and Applied Mathematics, 13,
  297, \dodoi{10.1002/cpa.3160130207}

\bibitem[{{Roe}(1981)}]{1981JCoPh..43..357R}
{Roe}, P.~L. 1981, Journal of Computational Physics, 43, 357,
  \dodoi{10.1016/0021-9991(81)90128-5}

\bibitem[{{Sano} {et~al.}(2013){Sano}, {Inoue}, \&
  {Nishihara}}]{2013PhRvL.111t5001S}
{Sano}, T., {Inoue}, T., \& {Nishihara}, K. 2013, Physical Review Letters, 111,
  205001, \dodoi{10.1103/PhysRevLett.111.205001}

\bibitem[{{Sano} {et~al.}(2012){Sano}, {Nishihara}, {Matsuoka}, \&
  {Inoue}}]{2012ApJ...758..126S}
{Sano}, T., {Nishihara}, K., {Matsuoka}, C., \& {Inoue}, T. 2012, \apj, 758,
  126, \dodoi{10.1088/0004-637X/758/2/126}

\bibitem[{{Schekochihin} {et~al.}(2004){Schekochihin}, {Cowley}, {Taylor},
  {Maron}, \& {McWilliams}}]{2004ApJ...612..276S}
{Schekochihin}, A.~A., {Cowley}, S.~C., {Taylor}, S.~F., {Maron}, J.~L., \&
  {McWilliams}, J.~C. 2004, \apj, 612, 276, \dodoi{10.1086/422547}

\bibitem[{{Shen} {et~al.}(2012){Shen}, {Zha}, \&
  {Huerta}}]{2012JCoPh.231.6233S}
{Shen}, Y., {Zha}, G., \& {Huerta}, M.~A. 2012, Journal of Computational
  Physics, 231, 6233, \dodoi{10.1016/j.jcp.2012.04.015}

\bibitem[{{Shima} \& {Kitamura}(2011)}]{2011AIAAJ..49.1693S}
{Shima}, E., \& {Kitamura}, K. 2011, AIAA Journal, 49, 1693,
  \dodoi{10.2514/1.J050905}

\bibitem[{{Shima} \& {Kitamura}(2013)}]{2013AIAAJ..51..992S}
---. 2013, AIAA Journal, 51, 992, \dodoi{10.2514/1.J052046}

\bibitem[{{Shu} \& {Osher}(1988)}]{1988JCoPh..77..439S}
{Shu}, C., \& {Osher}, S. 1988, Journal of Computational Physics, 77, 439,
  \dodoi{10.1016/0021-9991(88)90177-5}

\bibitem[{{Shu}(2009)}]{2009SIAMR..51...82S}
{Shu}, C.-W. 2009, SIAM Review, 51, 82, \dodoi{10.1137/070679065}

\bibitem[{{Steger} \& {Warming}(1981)}]{1981JCoPh..40..263S}
{Steger}, J.~L., \& {Warming}, R.~F. 1981, Journal of Computational Physics,
  40, 263, \dodoi{10.1016/0021-9991(81)90210-2}

\bibitem[{{Stone} {et~al.}(2008){Stone}, {Gardiner}, {Teuben}, {Hawley}, \&
  {Simon}}]{2008ApJS..178..137S}
{Stone}, J.~M., {Gardiner}, T.~A., {Teuben}, P., {Hawley}, J.~F., \& {Simon},
  J.~B. 2008, \apjs, 178, 137, \dodoi{10.1086/588755}

\bibitem[{{Toro} {et~al.}(1994){Toro}, {Spruce}, \&
  {Speares}}]{1994ShWav...4...25T}
{Toro}, E.~F., {Spruce}, M., \& {Speares}, W. 1994, Shock Waves, 4, 25,
  \dodoi{10.1007/BF01414629}

\bibitem[{{van Leer}(1979)}]{1979JCoPh..32..101V}
{van Leer}, B. 1979, Journal of Computational Physics, 32, 101,
  \dodoi{10.1016/0021-9991(79)90145-1}

\bibitem[{{Wada} \& {Liou}(1997)}]{1997SIAM.18..633K}
{Wada}, Y., \& {Liou}, M.-S. 1997, SIAM Journal on Scientific Computing, 18,
  633, \dodoi{10.1137/S1064827595287626}

\bibitem[{{Weiss} \& {Smith}(1995)}]{1995AIAAJ..33.2050W}
{Weiss}, J.~M., \& {Smith}, W.~A. 1995, AIAA Journal, 33, 2050,
  \dodoi{10.2514/3.12946}

\bibitem[{{Xisto} {et~al.}(2014){Xisto}, {P{\'a}scoa}, \&
  {Oliveira}}]{2014JCoPh.275..323X}
{Xisto}, C.~M., {P{\'a}scoa}, J.~C., \& {Oliveira}, P.~J. 2014, Journal of
  Computational Physics, 275, 323, \dodoi{10.1016/j.jcp.2014.07.009}

\bibitem[{{Zenitani} \& {Miyoshi}(2011)}]{2011PhPl...18b2105Z}
{Zenitani}, S., \& {Miyoshi}, T. 2011, Physics of Plasmas, 18, 022105,
  \dodoi{10.1063/1.3554655}

\end{thebibliography}

\clearpage
\gdef\thetable{\arabic{table}}
 \begin{threeparttable}[h!]
{\tiny
\caption{Dissipation coefficients in the mass flux of various schemes at subsonic range.}
\begin{tabular}{ccccc}
 \hline \hline
Scheme &$D^{(\rho)}$ &$D^{(u)}$ &$D^{(P)}$ &Comment \\ \hline
AUSM$^+$ &$|\bar{u}-\Delta M^4 a/8|$&$\bar{\rho} \bar{M} \bar{M^{2}}+\Delta \rho/2$& 0 & \\
SHUS &$|\bar{u}|$&$\bar{\rho} \bar{M}$&$(1-|\bar{M}|)/a$& \\
SLAU &$|u_{1/2}|$&0&$(1-|\bar{M}|)^2/a$& Not reduced to a one-side value at $|M|=1$.\\
HLL & $(1+|\Delta M|/2) a$& $\bar{\rho} \bar{M}/(1+|\Delta M|/2)$ &0& Not preserves a contact discontinuity. \\
HLLC &$|u_{1/2}|$& $\rho_L \rho_R M_{1/2} / \bar{\rho}$ & $(1-\rho_{\rm up} |M_{1/2}|/\bar{\rho})/a$& Coefficients are derived approximately.\\
Present &$|\bar{u}-\Delta M^4 a/8|$&$\bar{\rho} \bar{M} \bar{M^2}+\Delta \rho/2$& $(1-|M_{1/2}|)\rho_{\rm up}/\bar{\rho} a$ & \\
 \hline
\end{tabular}
\label{tab:massflux} 
\begin{tablenotes}
 \item[] Note: $\bar{X} = (X_L + X_R)/2$, $\Delta X=X_R-X_L$, $\rho_{\rm up} = \rho_L \; (M_{1/2}>0) \; {\rm or} \; \rho_R \; ({\rm else})$ 
\end{tablenotes}
}
 \end{threeparttable}

\clearpage
\gdef\thefigure{\arabic{figure}}

\end{document}